\documentclass[aps,prx,reprint, superscriptaddress, longbibliography]{revtex4-2}
\usepackage[section]{placeins}

\makeatletter
\def\input@path{{../}}
\makeatother

\usepackage[mode=image]{standalone}
\graphicspath{./}

\usepackage{lineno}
\usepackage{graphicx}  
\usepackage{dcolumn}   
\usepackage{bm}        
\usepackage{amssymb}   
\usepackage{amsmath,stmaryrd}
\usepackage{blkarray, multirow, graphicx, diagbox, color, colortbl}
\usepackage[dvipsnames]{xcolor}
\usepackage{bbm, bbold}
\usepackage{glossaries}
\usepackage{algorithm}
\usepackage{algpseudocode}
\usepackage{hyphenat}
\usepackage{ifthen}
\usepackage[colorlinks, linkcolor = blue, citecolor = blue, filecolor = black, urlcolor = blue]{hyperref}
\usepackage{dsfont}
\usepackage{xkeyval}
\usepackage{moreverb}
\usepackage{rotating}
\usepackage{wrapfig}
\usepackage{slashbox}
\usepackage{xspace}
\usepackage{nicefrac}
\usepackage[]{units}
\usepackage{physics}
\usepackage{booktabs}
\usepackage{braket}
\usepackage[inline]{enumitem}
\usepackage{tabto}
\usepackage{listings}
\usepackage[normalem]{ulem}
\usepackage{xstring}
\def\ReplaceStr#1{%
	\IfSubStr{#1}{p}{%
		\StrSubstitute{#1}{p}{.}}{#1}}
\usepackage[english]{babel}

\usepackage[caption=false]{subfig}
\usepackage[capitalise]{cleveref} %
\usepackage{chemformula}
\usepackage{algorithm}

\newcommand{\nodagger}[0]{{\vphantom{\dagger}}}

\newcommand{\updated}[1]{{\textcolor{black}{#1}}}

\newcommand{\vecbra}[1]{\langle\mspace{-3mu}\langle #1 |}
\newcommand{\vecket}[1]{| #1 \rangle\mspace{-3mu}\rangle}
\newcommand{\vecbraket}[2]{\langle\mspace{-3mu}\langle #1 | #2 \rangle\mspace{-3mu}\rangle}

\newcommand{\blackcircle}{%
  \tikz[baseline=-0.5ex]\filldraw[black] (0,0) circle (0.65ex);%
}

\newcommand{\blackcross}{%
  \tikz[baseline=-0.6ex, x=1ex, y=1ex]{%
    \draw[line width=0.25ex, black] 
      (0,-0.4) -- (1,0.6)
      (0,0.6) -- (1,-0.4);
  }%
}

\newacronym{OBC}{OBC}{open boundary condition}
\newacronym{PBC}{PBC}{periodic boundary condition}
\newacronym{DSP}{DSP}{dissipative state preparation}
\newacronym{MPS}{MPS}{matrix\hyp product state}
\newacronym{MPO}{MPO}{matrix\hyp product operator}
\newacronym{ED}{ED}{exact diagonalization}
\newacronym{QJ}{QJ}{quantum jumps}
\newacronym{BEC}{BEC}{Bose-Einstein condensate}
\newacronym{DPT}{DPT}{dissipative phase transition}
\newacronym{CDW}{CDW}{charge density wave}
\newacronym{1D}{1D}{one dimensional}
\newacronym{LSE-TDVP}{LSE-TDVP}{local subspace expansion time-dependent variational principle}
\newacronym{TDVP}{TDVP}{time-dependent variational principle}
\newacronym{ONB}{ONB}{orthonormal basis}
\newacronym{2TDVP}{2TDVP}{two-site time-dependent variational principle}
\newacronym{HEOM}{HEOM}{hierarchy of equations of motion}
\newacronym{DMRG}{DMRG}{density matrix renormalization group}
\newacronym{SL}{SL}{symmetrically\hyp localized}
\newacronym{TN}{TN}{tensor\hyp network}
\newacronym{CLIK-MPS}{CLIK-MPS}{complex-time Lindbladian Krylov subspace matrix\hyp product states}
\newacronym{LSE}{LSE}{local subspace expansion}
\newacronym{SVD}{SVD}{singular value decomposition}
\newacronym{QME}{QME}{quantum Mpemba effect}
\newacronym{KPZ}{KPZ}{Kardar\hyp Parisi\hyp Zhang}
\newacronym{OBDM}{OBDM}{one\hyp body reduced density matrix}    
\begin{document}
\author{Philipp Westhoff}\thanks{p.westhoff@physik.uni-muenchen.de}
\affiliation{Department of Physics, Arnold Sommerfeld Center for Theoretical Physics (ASC), Munich Center for Quantum Science and Technology (MCQST), Ludwig-Maximilians-Universit\"{a}t M\"{u}nchen, 80333 M\"{u}nchen, Germany}
\author{Mattia Moroder}\thanks{moroderm@tcd.ie}
\affiliation{School of Physics, Trinity College Dublin, College Green, Dublin 2, D02K8N4, Ireland}
\affiliation{Trinity Quantum Alliance, Unit 16, Trinity Technology and Enterprise Centre, Pearse Street, Dublin 2, D02YN67, Ireland}
\author{Ulrich Schollwöck}\thanks{schollwoeck@lmu.de}
\affiliation{Department of Physics, Arnold Sommerfeld Center for Theoretical Physics (ASC), Munich Center for Quantum Science and Technology (MCQST), Ludwig-Maximilians-Universit\"{a}t M\"{u}nchen, 80333 M\"{u}nchen, Germany}
\author{Sebastian Paeckel}\thanks{sebastian.paeckel@physik.uni-muenchen.de}
\affiliation{Department of Physics, Arnold Sommerfeld Center for Theoretical Physics (ASC), Munich Center for Quantum Science and Technology (MCQST), Ludwig-Maximilians-Universit\"{a}t M\"{u}nchen, 80333 M\"{u}nchen, Germany}

\def\thetitle{Tensor Network Framework for Lindbladian Spectra and Steady States}
\title{\thetitle}
\begin{abstract}
Quantum systems coupled to (non\hyp)Markovian environments attract increasing attention due to their peculiar physical properties.
Exciting prospects such as unconventional non\hyp equilibrium phases beyond the Mermin\hyp Wagner limit or dissipative state preparation
demand a systematic analysis of quantum many\hyp body phases out of equilibrium.
Akin to the equilibrium case, this requires the computation of the low\hyp lying eigenstates of Lindbladians, a problem challenging conventional approaches for simulating quantum many\hyp body systems.
Here, we undertake a first step to overcome this limitation and introduce a tensor\hyp network\hyp based framework to systematically compute not only steady states, but also low\hyp lying excited states for large, driven quantum many\hyp body systems.
Our framework is based on recent advances utilizing complex\hyp time Krylov spaces, and we leverage these ideas to create a toolbox tailored to solve the challenging non\hyp Hermitian eigenvalue problem ubiquitous in open quantum systems.
At the example of the interacting Bose\hyp Hubbard model driven by dissipation\hyp assisted hopping, we demonstrate the high efficiency and accuracy.
%
%
\updated{From a reliable finite-size scaling analysis of the spectral gap, we find strong evidence for nonlinear hydrodynamic behavior consistent with Kardar-Parisi-Zhang-type superdiffusive relaxation and establish the existence of exponentially accelerated, anomalous relaxation.}
This method unlocks the capability of spectral analysis of generic open quantum many\hyp body systems, suitable also for non\hyp Markovian environments.
\end{abstract}

\maketitle
\section{Introduction}
\label{sec:intro}
\begin{figure}[t!]
    \centering
    \subfloat[\label{subfig:complex:time:first:a}]{
        \includegraphics[width=0.8\columnwidth]{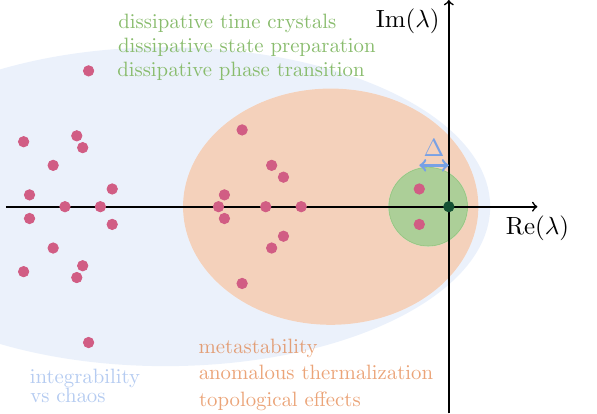} 
        }
        \vspace{-0.2cm} 
    \subfloat[\label{subfig:complex:time:first:b}]{
        \includegraphics[width=0.9\columnwidth]{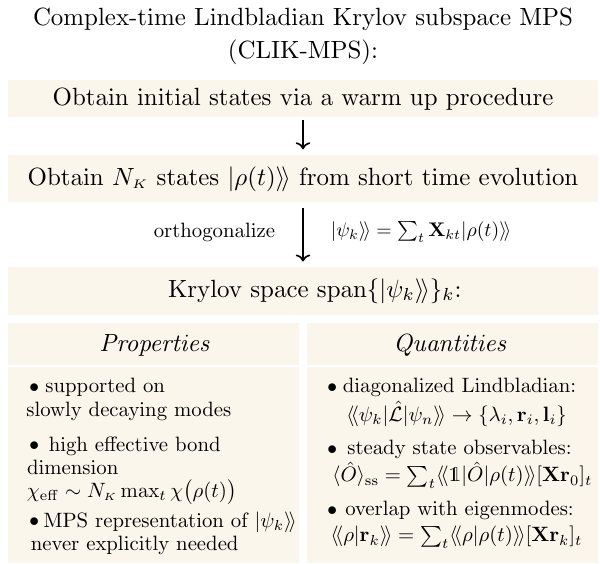}
        }
        \vspace{-0.2cm} 
    \subfloat[\label{subfig:hilbert:space:dim}]{
    \begin{tabular}{||c|c|c||}
    \hline
    Method & Target quantity & $\mathrm{dim}\, \mathcal{H}$
    \\ \hline\hline
    VQA \cite{yoshioka2020,Liu2021, Santos2025} & steady state & $2^{12}\approx 10^3$
    \\ \hline
    NQS \cite{Hartmann2019, Vicentini2019, mellak2024deep} & steady state &  $2^{16} \approx 10^5$ 
    \\ \hline
    Variational MPO \cite{Prosen2009, Cui2015, Kshetrimayum2017, Guo2022} & steady state & $2^{100} \approx 10^{30}$*
    \\ \hline
    Quantum MC \cite{Nagy2020quantum, Xu2023quantumkineticmontecarlo} & steady state & $2^{40} \approx 10^{12}$
    \\ \hline 
    \updated{MPS time evolution \cite{Zhan2025}} & \updated{steady state, gap} & \updated{$2^{40}\approx 10^{12}$
}
    \\ \hline 
    ED \cite{Almeida2025, Richter2025} & full spectrum & $2^{12}\approx 10^3$
    \\ \hline
    \acrshort{CLIK-MPS} & steady state, gap & \updated{$2^{176} \approx 10^{53}$}
    \\ \hline
    \acrshort{CLIK-MPS} & spectral region & $2^{70} \approx 10^{21}$
    \\ \hline
    \end{tabular}
        }
    \caption{
       Panel a): physical relevance of different parts of the Lindbladian spectrum.
        In general, the analysis of many\hyp body Markovian open quantum systems relies on the Lindbladian steady state (dark green), the Liouvillian gap $\Delta$ (green, blue), a set of low\hyp lying eigenvalues (orange), or the full spectrum (blue).
        Panel b): a schematic depiction of \gls{CLIK-MPS}.
        Panel c): maximal Hilbert space dimensions and accessible Lindbladian spectral regions for various existing methods and comparison to our new framework~\gls{CLIK-MPS}.
        The * indicates that for variational \gls{MPO} methods, large system sizes of about $100$ spins can be reached only for local Lindbladians that satisfy entanglement area laws \cite{Prosen2009}.
        }
    \label{fig:complex:time:first}
\end{figure}

In recent years, the rapid experimental development of quantum simulators~\cite{Altman2021, Gyger2024}, digital quantum processors~\cite{Madsen2022, Acharya2025}, the control over strong light\hyp matter interactions~\cite{Baranov2018, Qin2024}, as well as the unprecedented precision in optically driving low\hyp dimensional materials~\cite{Zribi2019, Duan2021} put modern quantum physics at the brink of experimentally utilizing strong correlations in many\hyp body systems to overcome classical limitations in terms of computational, algorithmic and metrological capabilities.
Alongside this remarkable progress, the practical necessity to theoretically describe and analyze open quantum systems became increasingly pressing.
Understanding the intriguing physics of quantum many\hyp body systems coupled to large environments poses mathematical challenges, which reach far beyond the well\hyp known frameworks for analyzing and classifying closed systems.
Already for the simplest example of Markovian environments, analytic treatments are in general limited to non\hyp interacting systems via third quantization \cite{Prosen2008} or the Lyapunov equation \cite{Landi2022}.
As a consequence, there is an urgent need for an efficient and flexible framework to overcome these limitations, and accessing information about Lindbladian spectra plays a crucial role in achieving this goal.
Methodological developments so far have been limited to compute the steady state, i.e., the open\hyp system analogue of the ground state in closed systems, with approaches ranging from~\gls{TN} methods~\cite{Verstraete2004,Prosen2009,Cui2015,Mascarenhas2015, Werner2016,Kshetrimayum2017,Gangat2017,Somoza2019,Wolff2020, Guo2022}, potentially combined with Monte\hyp Carlo sampling~\cite{Pichler2013,Sarkar2014,Moroder2022,Xie2024, Hryniuk2024}, to neural quantum states~\cite{Hartmann2019,Vicentini2019,mellak2024deep}, quantum algorithms~\cite{Cleve2019,yoshioka2020,Liu2021, Ding2024, Santos2025, Xie_2025} and phase space methods~\cite{Tosca2025}.
However, besides the steady state, the capability to determine the low\hyp lying eigenstates has become increasingly relevant.
For instance, the dissipative gap, i.e. the real part of the first excited eigenvalue, dictates the experimentally relevant relaxation timescale of the system~\cite{Carollo2021,westhoff2025fast}.
Many other phenomena, such as dissipative phase transitions \cite{Minganti2018, Minganti2023}, anomalous thermalization processes \cite{Carollo2021, Moroder2024}, topological effects \cite{Song2019, Lieu2020, Yoshida2020} and metastability \cite{Macieszczak2016}, require the explicit knowledge of the Lindbladian spectrum and eigenvectors. 
In~\cref{subfig:complex:time:first:a} we summarize the relations between physical phenomena and the corresponding parts of the Lindbladian spectrum.
Compared to the isolated case characterized by a Hamiltonian, diagonalizing the Lindbladian $\hat{\mathcal L}$ presents two main additional challenges.
First, $\hat{\mathcal L}$ is non\hyp hermitian, which prevents the direct application of variational algorithms such as the \gls{DMRG}~\cite{White1992,White1993,Schollwoeck2011}.
\updated{Here, it is important to emphasize that while there are methods to efficiently compute the steady state, or the spectral gap~\cite{Chan2005,Vanderstraeten2016,Zhong2025,Zhan2025}, these approaches typically do not efficiently generalize to the precise determination of several low\hyp lying left/right eigenstates of the Lindbladian.}
Second, its dimensionality grows quadratically faster than the Hamiltonian's upon increasing the system size, which strongly limits the sizes of systems amenable to an \gls{ED} treatment.
\updated{Taking a conceptual point of view, it is peculiar that previous approaches rarely took advantage of the additional spectral structures inherent to Lindbladians.
For instance, eigenstates always come in complex\hyp conjugated pairs and the spectrum is restricted to the semi\hyp infinite half space $\operatorname{Re}(\lambda)\leq 0$.
For that purpose, we propose \acrfull{CLIK-MPS}, a \gls{TN}\hyp based framework that allows us to specifically target a set of low\hyp lying Lindblad eigenvalues and eigenmodes with very high accuracy and efficiency, exploiting exactly this additional structure of the Lindbladian.
}
It is built upon the observation that the short\hyp time dynamics, generated by $\hat{\mathcal L}$ acting on general initial states (i.e., vectorized density operators $\vecket{\rho}$) can be simulated efficiently using~\gls{TN} methods due to the inherent non\hyp Hermiticity of the Lindbladian~\cite{Minganti2022}.
Physically, this can be understood by noting that, in general, dissipation tends to localize the physical degrees of freedom, an effect that has been observed in several numerical studies, and which suppresses entanglement growth in the dynamics~\cite{Moroder2022,Preisser2023,caceffo2024fateentanglement}.
We then utilize the set of time\hyp evolved density operators $\left\{\vecket{\rho(t_n)}\right\}_n$ ($n\in\mathds N$) to construct a basis for a Krylov subspace generated by $\mathrm{exp}(\hat{\mathcal L} t_n)$ and we show that the subspace overlap of this basis with the subspace spanned by the low\hyp lying eigenmodes of $\hat{\mathcal L}$ can be enhanced significantly by choosing proper complex time contours~\cite{Grundner2024, Paeckel2024}.
Upon \updated{expoiting the trace preserving property of the Lindbladian and} introducing a physically motivated warmup\hyp procedure to create rapidly converging initial states, we show that the Lindbladian can be approximated efficiently in this basis with the low\hyp lying eigenstates exhibiting an unprecedented precision.
We demonstrate the efficiency of our method by studying a many\hyp body dissipative bosonic model.
As a key feature, this model exhibits a non\hyp trivial spectral density in momentum space.
Consequently, the Lindblad operators realize a minimal model for a coupling between physical single\hyp particle states and a non\hyp Markovian environment, i.e., a Markovian embedding~\cite{Moroder2022,Xie2024}.
Our simulations therefore represent a testcase for determining the low\hyp lying eigenstates of a system coupled to a non\hyp Markovian environment, reaching striking Hilbert space dimensions (see \cref{subfig:hilbert:space:dim}), and achieving very good accuracy both for the spectrum and the eigenstates.
The article is organized as follows: In~\cref{sec:clik-mps}, we introduce the \acrshort{CLIK-MPS} framework, focusing on the role of the complex-time evolution, optimal initialization, and the efficient computation of observables in the Krylov subspace.
Then, in~\cref{sec:results}, we apply it to the dissipative Bose–Hubbard model, analyzing its steady state, dissipative gap, and anomalous thermalization behavior.
Finally, in~\cref{sec:conclusion}, we summarize our findings and outline promising directions for future research.
\section{\Acrfull{CLIK-MPS}}
\label{sec:clik-mps}
In this work, we introduce a framework based on~\gls{TN} methods which allows to reliably and efficiently compute the low\hyp lying eigenstates of a quantum system weakly coupled to a Markovian (i.e. memoryless) environment.
These kinds of systems are described by a Lindblad master equation \cite{Lindblad1976}
\begin{equation}
    \frac{\mathrm{d} \hat{\rho}(t)}{\mathrm{d}t} = \mathcal{L} \hat{\rho}(t) = -\mathrm i [\hat{H}, \hat{\rho}(t)] + \sum_l \Big(\hat{L}^\nodagger_l \hat{\rho}(t) \hat{L}^\dagger_l - \frac{1}{2} \{ \hat{L}^\dagger_l \hat{L}^\nodagger_l, \hat{\rho}(t)\}\Big) \;,
    \label{eq:Lindblad}
\end{equation}
where $[\cdot, \cdot]$ and $\{\cdot, \cdot\}$ indicate the commutator and the anticommutator, respectively, $\hat{\rho}$ is the system's density matrix, $\hat{H}$ is the Hamiltonian and the influence of the environment on the system is captured by the so\hyp called jump operators $\hat{L}_l$.
Importantly, using techniques such as Markovian embeddings, this form also allows to describe non\hyp Markovian environments with non\hyp trivial, structured spectral densities~\cite{Moroder2022,Xie2024}.
To study the spectral properties of the Lindbladian, we make use of vectorization, which consists of mapping density matrices to vectors $\hat{\rho} \to \vecket{\rho}$ and the Lindbladian superoperators to operators $\mathcal{L} \to \mathcal{\hat{L}}$.
Following the procedure outlined for instance in Ref.~\cite{Landi2022}, the vectorized Lindbladian can be written as
\begin{equation}
    \begin{aligned}
       &\hat{\mathcal{L}} = -\mathrm i \hat{H} \otimes  \hat{\mathds{1}} +   \hat{\mathds{1}} \otimes \mathrm i\hat{H}^{\scriptscriptstyle T} + \\
       &\sum_l \hat{L}^{\nodagger}_l \otimes \left ( \hat{L}^{\dagger}_l \right ) ^{\scriptscriptstyle T} -\frac{1}{2} \hat{L}^{\dagger}_l \hat{L}^{\nodagger}_l \otimes  \hat{\mathds{1}}  - \frac{1}{2}  \hat{\mathds{1}} \otimes \left( \hat{L}^{\dagger}_l \hat{L}^{\nodagger}_l \right) ^{\scriptscriptstyle T}.
    \end{aligned}
    \label{eq:Lindblad:vectorized}
\end{equation}
For two operators $\hat \rho, \hat \mu$ given in the vectorized form $\vecket{\rho}, \vecket{\mu}$, we define a generalized scalar product relating the two representations
\begin{equation}
     \operatorname{Tr}\left( \hat\rho^\dagger \hat\mu \right) = \vecbraket{\rho}{\mu} \; .
\end{equation}
\updated{
We always represent states and operators in the~\gls{MPS} and~\gls{MPO} representation \cite{White1992,Schollwoeck2011}, but the framework is of course independent of the choice of the~\gls{TN} representation.
Density matrices are vectorized by doubling the system size, alternating physical and auxiliary sites avoiding the introcution of long\hyp range couplings~\cite{Casagrande2021} in the \gls{MPO} representation of the vectorized Lindbladian \cref{eq:Lindblad:vectorized}~\cite{Prosen2009, Wolff2020, Somoza2019}.
}
In this section, we detail the main components of~\gls{CLIK-MPS} to solve the eigenvalue problem of~\cref{eq:Lindblad:vectorized} for low\hyp lying eigenvalues and eigenmodes.
We begin with a short survey introducing the main ideas of the framework.
Thereafter, we discuss some key aspects where we put special emphasis on the tilted, complex contour to target the low\hyp lying eigenmodes of $\hat{\mathcal L}$ with high precision, as well as the construction of tailored initial states, which are necessary to achieve fast convergence.
Finally, we demonstrate how expectation values can be evaluated efficiently and show how symmetries of the Lindbladian can be used to increase the precision of the computation of expectation values in the steady state.
\subsection{Survey of~\gls{CLIK-MPS}}
\begin{figure}
    \centering
    \includegraphics[width=0.98\columnwidth]{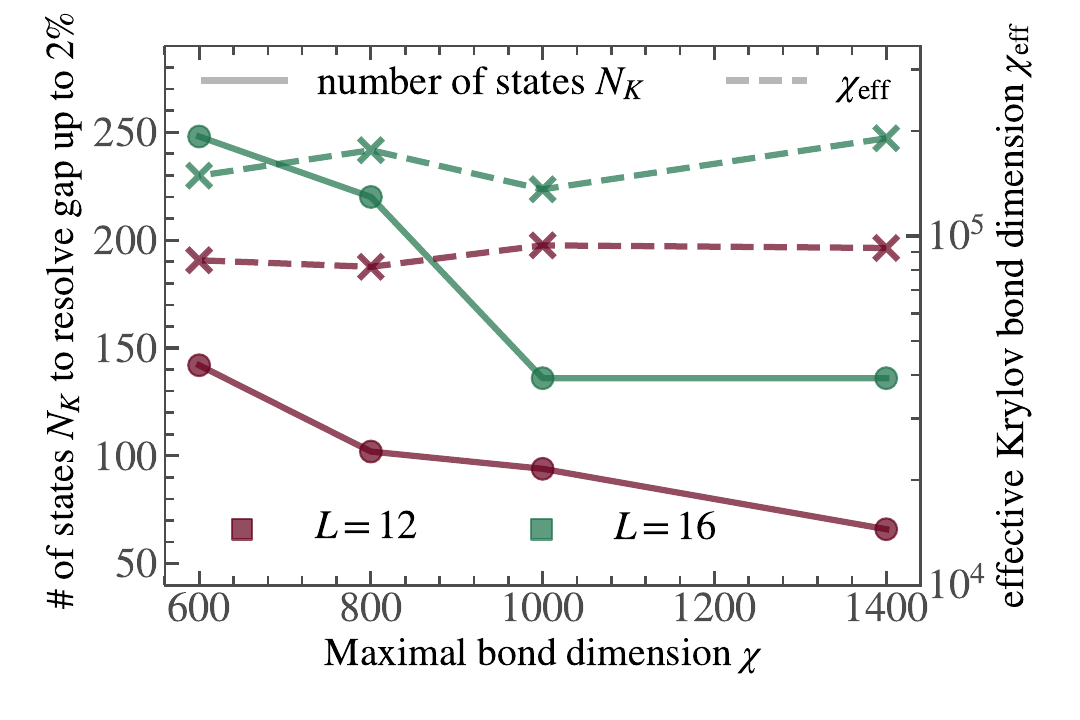}
    \caption{\updated{The role of the effective bond dimension. Left: Number of time evolved states $N_{\scriptscriptstyle K}$ needed to resolve the dissipative gap ($\Delta=0.293, \, 0.174$ for $L=12 , \, 16$, respectively) up to a relative error of $2\%$ for different system sizes.  $N_{\scriptscriptstyle K}$ decreases upon increasing the maximal bond dimension $\chi$ during time evolution. 
    Right: the effective bond dimension $\chi_\mathrm{eff}$ defined in~\cref{eq:eff:bdim:simple} is independent of $\chi$, showing that one can trade a lower bond dimension during time evolution for a larger number of linearly independent time-evolved vectors.
    For all calculations we chose $J=1$, $\kappa=2$ and $U=0$, as well as $\delta t = 0.2$.
    }}
    \label{fig:effective:bond:dim}
\end{figure}

Approximating the eigenvalue problem of high\hyp dimensional non\hyp Hermitian operators in a low\hyp dimensional subspace is a particularly challenging task for~\gls{TN}\hyp based methods.
The absence of Hermiticity prohibits the formulation in terms of a Rayleigh\hyp Ritz variational principle and attempts to solve equivalent hermitized problems exhibit a significant increase of computational complexity, mostly due to the artificial introduction of long\hyp ranged couplings and degraded convergence~\cite{Cui2015,Guo2022}.
However, for general Lindbladians it is well known that the steady state, i.e. the normalized right eigenmatrix of $\hat{\mathcal L}$ corresponding to the extremal eigenvalue $\lambda_1 = 0$, can be approximated via a time\hyp evolution, where the convergence rate is exponential in the Lindbladian gap w.r.t. the evolution time.
This can be seen directly by expanding the dynamics of an arbitrary initial density operator $\vecket{\rho_0}$ in the eigenbasis of the Lindbladian
\begin{equation}
    \vecket{\rho(t)}=\mathrm e^{\hat{\mathcal L}t}\vecket{\rho_0}= \vecket{\rho_\mathrm{ss}}+\sum_{k=2}^{D^2}\mathrm e^{\lambda_k t}\vecbraket{l_k}{\rho_0}\vecket{r_k} \;,
    \label{eq:lindblad:decomposition}
\end{equation}
where $\lambda_k$ are the eigenvalues of $\hat{\mathcal L}$, $\vecket{r_k}$ ($\vecket{l_k}$) the corresponding right (left) eigenvectors, and $D=\operatorname{dim} \mathcal H$ denotes the dimension of the Hilbert space of the underlying physical system.
Here, the contribution from every eigenmode decays exponentially in time with a rate dictated by $\mathrm{Re}(\lambda_k)$, except for the steady state $\vecket{\rho_\mathrm{ss}}$.
\updated{Note that \cref{eq:lindblad:decomposition} holds only when evolving a density matrix. 
If one considers a \textit{traceless} initial state $\vecket{\rho_0}$, the steady state component is suppressed owing to the trace preservation, and the time evolution converges to the slowest right eigenvector $\vecket{r_2}$, instead.
We will thus use such states later on to resolve the low-lying eigenvalues beyond the steady state.}
We emphasize that the complex eigenvalues of~\cref{eq:Lindblad:vectorized} satisfy $\operatorname{Re} \lambda_k \leq 0$.
This time\hyp evolution can be computed efficiently using~\gls{TN} methods as well as a vectorized representation of the Lindbladian and the density operators, yet evolution times are typically limited by the entanglement growth.
\label{subsec:survey}
\begin{figure}[t!]
    \centering
    \subfloat[\label{fig:complex:contour}]{
        \includegraphics[width=0.9\columnwidth]{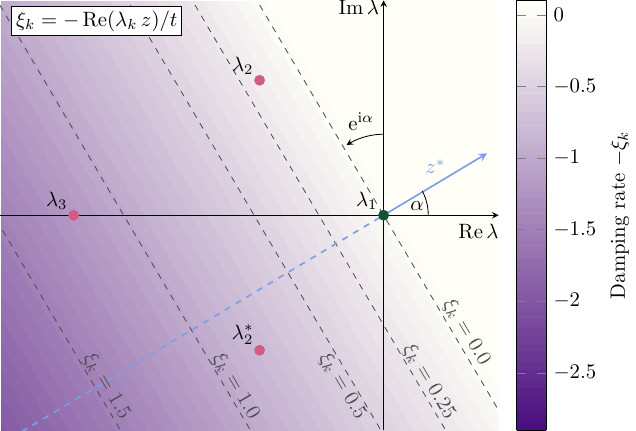}
    }%
    \vspace{-0.2cm}
    \subfloat[\label{fig:spec:random:states}]{
        \includegraphics[width=0.95\columnwidth]
        {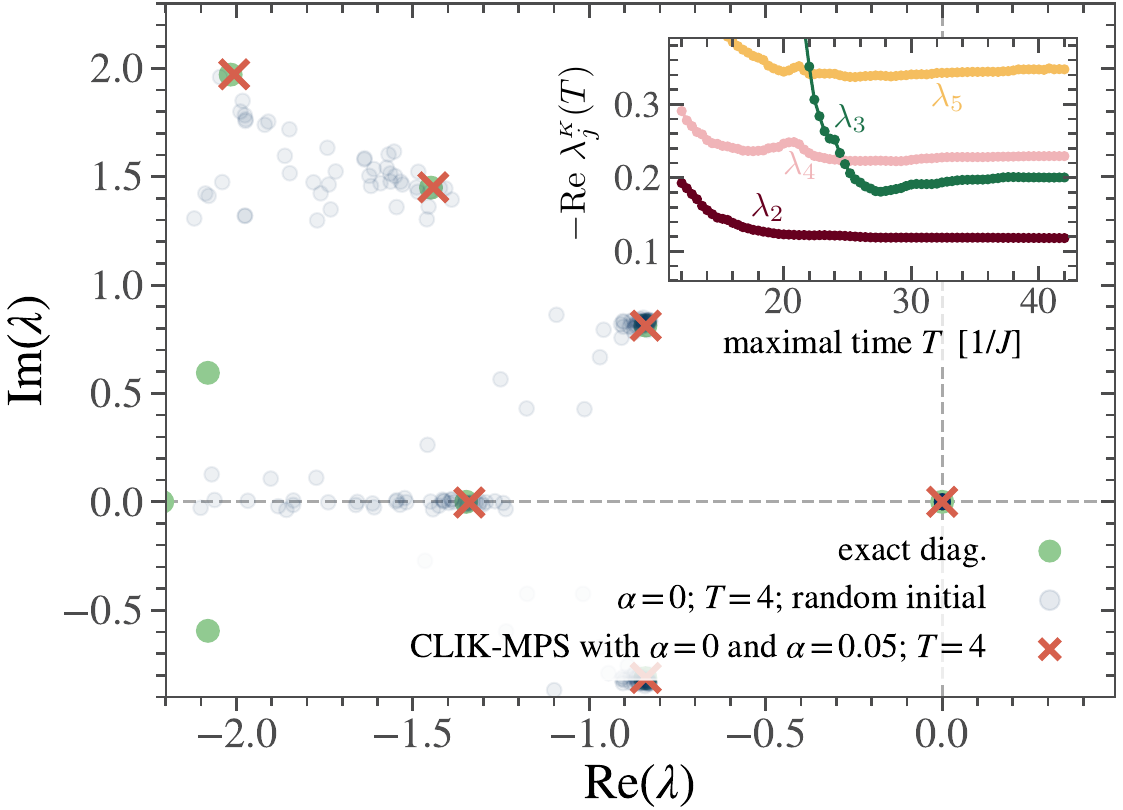}
    }
    \caption{
    Panel a): Linearly inclined complex time contours $z=t \mathrm e^{-i\alpha}$ alter the damping rates $\xi_k$ of Lindbladian eigenmodes (see~\cref{eq:lambda:complex:angle}).
    The color shading indicates the damping rate $\xi_k = -\operatorname{Re}(\lambda_k \,z)/t$ for an eigenvalue located at $\lambda_k=\operatorname{Re} \lambda_k + \mathrm i\operatorname{Im} \lambda_k$, when evolving along the contour $z$.
    Note the tilt of the contour lines $\xi_k = \mathrm{const.}$, which are orthogonal to $z^*$ (solid blue line).
    In the upper half plane ($\operatorname{Im} z > 0$) and for $\alpha>0$, excited, i.e., fast\hyp decaying, modes are enhanced by reducing their damping rate (brighter areas), while eigenmodes with eigenvalues on the lower half plane ($\operatorname{Im} z < 0$) are suppressed due to an increased damping rate (darker areas).
    Panel b): Impact of the initial state on the accuracy of the spectrum of the dissipative Bose\hyp Hubbard model, specified by~\cref{eq:model:hamiltonian,eq:model:jump-ops}, for $L=5$ and $N=4$. We compare \gls{ED} (\protect\blackcircle\updated{, green}) data to the spectrum from a naive Krylov method with a single time evolution (\protect\blackcircle\updated{, light blue}), with a randomly sampled initial state. In total, 50 initial states were generated. 
    The darkness of the blue color indicates how many points overlap at the respective position.
    Lastly, the results from \gls{CLIK-MPS} are shown (\protect\blackcross).
    Inset: First 4 slow decaying modes approximated using \gls{CLIK-MPS} for $L=20$, $N=10$ sites at $U/J=0.5$ and $\kappa/J=2$ depending on the maximal evolved time $T$.}
\end{figure}
The crucial observation is that the density operators at short and intermediate times already contain valuable information about the spectrum of the Lindbladian.
In~\gls{CLIK-MPS}, we use a set of $N_{\scriptscriptstyle K}$ time\hyp evolved density\hyp operators $\left\{\mathrm{exp}(\hat{\mathcal L} t_n)\vecket{\rho_0}\right\}$ where $t_n = n \delta t$ for some time step $\delta t>0$ to generate a basis for the right Krylov space of the operator exponential of the Lindbladian. \updated{Here, the subscript $K$ denotes that this number is related to the Krylov space.}
While \cref{eq:lindblad:decomposition} shows that contributions from highly excited states decay substantially faster than those from low-lying excited states, the latter contributions are generally also small, so the generated Krylov space can accurately represent only the steady state.
To solve this problem and generate a Krylov space that is well-suited to approximate also a set of low-lying Lindbladian right eigenmodes, we refine the naive time evolution in two ways:
\begin{enumerate}
    \item We consider time\hyp evolution contours in the complex plane, i.e., we take $\delta t\rightarrow \delta z\in\mathds C$, and we show that these complex time\hyp steps can be choosen in a way to optimize the convergence of specific eigenmodes by reducing their damping $\xi_k$ w.r.t. the contour, which is illustrated in~\cref{fig:complex:contour}.
    \item We introduce a warmup procedure to cheaply construct initial operators $\vecket{\rho_0}$, which are optimized to enhance the convergence of low\hyp lying excited states of the Lindbladian.
\end{enumerate}
In~\cref{fig:spec:random:states}, we demonstrate the improvements of~\gls{CLIK-MPS} over conventional time evolution for a dissipative system of interacting bosons with $L=5$ sites and $N=4$ particles (see~\cref{sec:results} for the model's details).
The strong dependence on the chosen initial states is demonstrated by the transparent dots corresponding to the eigenvalue approximations obtained from randomly chosen initial density operators and using conventional time evolution.
Note how these points are vastly scattered around the excited states obtained from exact diagonalization (green dots).
In turn, the crosses are generated using~\gls{CLIK-MPS}.
The first five eigenvalues are reproduced with excellent precision using the same maximum evolution time $T = 4$ as in the conventional time evolution.
For illustrative purposes, in the inset we display the convergence of the five lowest Lindbladian eigenvalues for a significantly larger system with $L=20$ sites and $N=10$ interacting bosons, using~\gls{CLIK-MPS}.
The resulting Hilbert space dimensions (c.f.~\cref{subfig:hilbert:space:dim}) have long been inaccessible to existing methods, particularly for computing excited eigenmodes.
The final ingredient of~\gls{CLIK-MPS} is the efficient representation of operators in the complex\hyp time Krylov subspace.
Clearly, constructing explicit~\gls{TN} representations of the orthogonalized basis states from the time\hyp evolved states would be prohibitively expensive, due to the fast increase of bond dimension when superimposing and orthogonalizing the~\gls{TN} representations of the vectorized density operators.
We circumvent this problem by computing the Gram matrix $\mathbf{M}_{ij} = \vecbraket{\rho(t_i)}{\rho(t_j)}$ of the evolved density operators and diagonalization $\mathbf{M}=\mathbf{U}\mathbf S\mathbf{U}^\dagger$, which is a computationally cheap operation.
Then, a transformation into an \gls{ONB} is given by \updated{$\mathbf{X} = \mathbf{U} \mathbf S^{-\nicefrac{1}{2}}$}, where additionally a deflation of linear dependent states can be performed by discarding small eigenvalues below the numerical precision~\cite{Paeckel2024}.
Numerical instabilities in this expression are discussed in detail in \cref{subsec:subspace-expansion} and alleviated by the methods described there.
Any superoperator $\hat{\mathcal O}$ is then easily represented in the Krylov space by computing its expectation values with all time\hyp evolved density operators $\mathbf O_{ij}=\vecbra{\rho(t_i)}\hat{\mathcal O}\vecket{\rho(t_j)}$, a task that can be parallelized trivially, and using the basis transformation constructed from the Gram matrix $\mathbf O^\mathrm{eff} = \mathbf X^\dagger \mathbf O \mathbf X$.
It is important to note that using this approach, we effectively describe density operators with a much higher bond dimension than the one actually used during the numerics~\footnote{In this work, we always refer to the bond dimension of a site tensor as the largest dimension of its set of virtual legs.}.
To illustrate this point, let $\chi(\vecket{\rho(t_i)}) = \chi_i$ be the bond dimension of the time\hyp evolved density operator at time step $t_i$ and denote by $N_{\scriptscriptstyle{K}}$ the number of evolved time steps, i.e., the number of (possibly linear dependent) basis states of the Krylov space $\mathcal K$.
If we would construct the~\gls{MPS} representation of the orthonormal basis states of $\mathcal K$ explicitly, this would involve superpositions of the $N_{\scriptscriptstyle{K}}$ states $\left\{\vecket{\rho(t_i)}\right\}$ such that, to a good approximation, the bond dimension of the basis states of $\mathcal K$ scales as
\begin{equation}
\label{eq:eff:bdim:simple}
    \chi_\mathrm{eff} \sim N_{\scriptscriptstyle{K}} \max_{i} \chi_i \; .
\end{equation}
Later, we will construct Krylov spaces with $N_{\scriptscriptstyle{K}} \sim \mathcal O(100)$, i.e., we approximate the Lindbladian in a Krylov space whose basis states would be represented by~\gls{MPS} with bond dimensions exceeding those of the time\hyp evolved states by two orders of magnitude!
This provides an intuitive explanation for the astonishing precision of~\gls{CLIK-MPS} we observed during our simulations for the steady state as well as the low\hyp lying eigenvalues and eigenmodes of the Lindbladian.
A further interesting conclusion can be drawn by assuming that the actual parameter controlling the approximation quality is $\chi_\mathrm{eff}$.
In that case, the computational costs can be reduced significantly by reducing the maximum bond dimension during the time evolution, say by a factor of $\nu$ and in turn increasing the number of time steps by the same factor.
Owing to the cubic scaling of the computational costs on the~\gls{MPS} bond dimension, a speed up by a factor of $\nu^2$ can be expected.
However, as discussed in~\cref{app:sec:numerical:details}, care must be taken to avoid too small values of the \updated{maximal} bond dimension since otherwise numerical instabilities can occur.
\updated{In \cref{fig:effective:bond:dim} we analyze this argument quantitatively by calculating the dissipative gap of a Bose\hyp Hubbard system (c.f. \cref{sec:results}) for different system sizes. 
We vary the maximally allowed bond dimension $\chi$ during time evolution, which was always saturated. 
Then we track the Krylov space dimension $N_{\scriptscriptstyle K}$ necessary to achieve a fixed accuracy of 2\% in the resolution of the dissipative gap $-\mathrm{Re} \, \lambda_2$.
In agreement with our intuition, the number of time evolved states $N_{\scriptscriptstyle K}$ decreases with increasing bond dimension $\chi$.
We also find $\chi_\mathrm{eff}$ (see~\cref{eq:eff:bdim:simple}) to be approximately constant upon varying $\chi$, indicating that it is indeed a central parameter determining the computation's accuracy. Note that this argument is only valid in a bond dimension window: If the bond dimension is much to small, the time evolution completely misses the real dynamics and convergence is never reached.
}
%
%
\subsection{Constructing the Krylov subspace} 
\label{subsec:krylov:space}
\updated{
\gls{CLIK-MPS} is based on time evolutions of vectorized density operators by simulating the dissipative Lindbladian dynamics, i.e., solving a Schrödinger equation with a non\hyp Hermitian generator on a doubled Hilbert space.}
\updated{
The time evolution is performed using the single\hyp site variant of the~\gls{TDVP}~\cite{Haegeman2011,Paeckel2019}.
\Gls{TDVP} reformulates the problem of solving the equation of motion for the density operator as a whole~\cref{eq:Lindblad} into a local equation of motion for each site tensor of the vectorized density operator.
While this already simplifies the complexity drastically, further optimizations are in order to obtain a robust time stepper.
First, for our test system we choose the single\hyp site variant of~\gls{TDVP} due to the large local Hilbert space dimension of the bosons.
Because in its standard formulation this does not allow for an increase of the~\gls{MPS} bond dimension, we additionally use the~\gls{LSE}~\cite{Yang2020,Grundner2023} to grow the bond dimensions after solving the local equation of motion.
Second, the non\hyp Hermitian nature of the generator rules out the standard Lanczos\hyp based integrator.
Instead, we build on previous works exploring different strategies to address this issue~\cite{Moroder_diss}.
Since we are using a comparable small time step $\mathrm d t = 0.01$ for all large scale calculations, a straightforward Taylor expansion proved to be both, conceptually simple and numerically robust throughout the different parameter regimes.
We discuss the numerical implementation in detail in \cref{app:sec:numerical:details}.
}
The most naive scheme to construct a Krylov space is to perform a time evolution of an initial state $\vecket{\rho_0}$ where the dynamics is governed by $\hat{\mathcal L}$ (c.f.~\cref{eq:lindblad:decomposition}), and to store $N_{\scriptscriptstyle K} \in \mathds N$ time\hyp evolved states $\vecket{\rho(t_i)}$.
\cref{eq:lindblad:decomposition} clearly suggests that these states are a particularly good choice for spanning a Krylov subspace with a large overlap with the steady state of $\hat{\mathcal L}$~\cite{Minganti2022}.
We denote this subspace as $\mathcal{K}_0(\rho_0) = \mathrm{span}\big\{\vecket{\rho_0}, \; \mathrm{exp}(\hat{\mathcal{L}}\,\updated{\delta} t)\vecket{\rho_0}, \; \dots  \; (\mathrm{exp}(\hat{\mathcal{L}}\,\updated{\delta} t))^{N_{\scriptscriptstyle K}-1}\vecket{\rho_0}  \big\}$.
However, when aiming for a high approximation quality of low\hyp lying excited states, the exponential decay constitutes a major obstacle, because it quickly suppresses the desired eigenmodes during the time evolution.
We overcome this limitation by extending the time\hyp evolution contour into the complex plane. 
Let us denote the path along which the time evolution is performed by $z(t)\in\mathds C$, where $t = \lvert z \rvert$ is used to parametrize the path along the complex plane.
The goal is to adjust the exponential damping of some low\hyp lying eigenmodes; from~\cref{eq:lindblad:decomposition} we find that it is governed by
\begin{equation}
    \left\lvert\vecbraket{l^\nodagger_k } {\rho(z)}\right\rvert
    \sim
    \mathrm e^{\operatorname{Re} z\lambda_k}
    \equiv
    \mathrm e^{-\xi_k t} \;. \label{eq:lambda:damping-rate}
\end{equation}
Here, \updated{$\xi_k = -\operatorname{Re} (z \lambda_k) / t$} denotes the damping rate of the $k$\hyp th eigenmode.
Let us now consider the linearly inclined complex\hyp time contours $z(t) = t \mathrm e^{-\mathrm i \alpha}$ with $\alpha > 0$ as shown in~\cref{fig:complex:contour}.
Evaluating the dynamics along this contour, the damping constants of the excited states are given by
\begin{equation}
    \xi_k = -\operatorname{Re}(\lambda_k)\cos\alpha - \operatorname{Im}(\lambda_k)\sin\alpha \, .
    \label{eq:lambda:complex:angle}
\end{equation}
We observe that the effect of the tilted contour is to increase the damping of eigenmodes with $\operatorname{Im}\lambda_k<0$, while reducing the damping for eigenmodes with $\operatorname{Im}\lambda_k>0$ and thereby enhancing their contribution to the dynamics, i.e., the Krylov space.
From a geometric point of view,
this can be understood by noting that the lines of constant damping rate $\xi$ 
are orthogonal to the path followed by the complex conjugated contour $z^*(t)$, as shown in~\cref{fig:complex:contour}.
In the absence of a tilt $\alpha = 0$, these equi\hyp damping lines are parallel to the imaginary axis of the complex plane.
For any finite tilt angle $\alpha > 0$, these lines are rotated counterclockwise and therefore slow down the damping of excited eigenmodes, if $\operatorname{Im} \lambda_k > 0$.
Crucially, because the Lindbladian modes come in complex conjugate pairs, it suffices to accurately determine, for instance, those eigenmodes $\vecket{r_k}$ with a positive imaginary part.
A poor convergence of the eigenmodes with $\mathrm{Im}\, \lambda_k<0$ is no problem at all because they are readily obtained from the ones with a positive imaginary part via complex conjugation $\vecket{r_k^\nodagger} \rightarrow \vecket{r^\dagger_k}$, $\lambda_k \to \lambda_k^*$. 
\updated{
Note however, that $\alpha$ cannot be arbitrarily chosen. If $\alpha>\alpha_\mathrm{max}$ for some $\alpha_\mathrm{max}$ depending on the spectrum of $\hat{\mathcal L}$, bulk states with large imaginary part get exponentially enhanced during time evolution and the resulting Krylov space is instable and resolves the low-lying spectrum poorly. 
In \cref{app:sec:krylov:space} we provide detailed information on how to curtail these problems.
}
We now project the full Lindbladian onto the complex\hyp time Krylov subspace $\mathcal{K}_\alpha(\rho_\alpha) = \mathrm{span}\big\{\vecket{\rho_\alpha}, \; \mathrm{exp}(\hat{\mathcal{L}}\,\delta z)\vecket{\rho_\alpha}, \; \dots  \; (\mathrm{exp}(\hat{\mathcal{L}}\,\delta z))^{N_{\scriptscriptstyle K}-1}\vecket{\rho_\alpha} \big\}$ with some new initial state $\vecket{\rho_\alpha}$ \updated{and the complex time step $\delta z = \mathrm e^{-\mathrm i \alpha}\delta t$}.
This results in a dramatic increase of accuracy on one quarter of the complex plane, allowing us to resolve low\hyp lying excited states to a very high precision.
In order to span an optimal Krylov space for both, the steady as well as the low\hyp lying excited state, we thus combine a time evolution with zero complex angle with one featuring a finite angle and construct the complex\hyp time Krylov space in which we approximate the Lindbladian via $\mathcal K_{0, \alpha} = \mathcal K_0(\rho_0)\oplus \mathcal K_\alpha(\rho_\alpha)$. \updated{Note that we propose to chose $\rho_\alpha$ as a \textit{traceless} state, so that it converges to the slowest decaying mode, further increasing the resolution of the low-lying spectrum beyond the steady state.}
Here, it should be noted that regarding the approximation quality of the low\hyp lying eigenmodes, an additional benefit comes from the fact that now each eigenmode contributes to $\mathcal K_{0, \alpha}$ with two different exponential damping rates.
The left eigenmodes of $\hat{\mathcal L}$ in general are not accurately described in $\mathcal K_{0, \alpha}$.
However, this can be resolved by performing the time evolution generated by $\hat{\mathcal L}^\dagger$, which can be written in the eigenbasis of $\hat{\mathcal L}$ as 
\begin{equation}
    \vecket{\rho^*(z)}=\mathrm e^{\hat{\mathcal L}^\dagger z}\vecket{\rho_0}= \sum_{k=1}^{D^2}\mathrm e^{\lambda_k^* z}\vecbraket{r_k}{\rho_0}\vecket{l_k} \;,
    \label{eq:lindblad:dagger:decomposition}
\end{equation}
where $\rho^*$ indicates that the dynamics is generated by $\hat{\mathcal L}^\dagger$.
The resulting Krylov space $\mathcal K^*_\alpha(\rho_0)$ spanned by these time-evolved states now includes contributions mainly from the slowly decaying left eigenmodes. Again, we utilize two time evolutions and build the Krylov space $\mathcal K_{0, \alpha}^* = \mathcal K_0^*(\rho_0)\oplus \mathcal K_\alpha^*(\rho_\alpha)$.
Importantly, for every Lindbladian, the left eigenspace corresponding to $\lambda_1=0$ features one eigenmatrix $\vecket{l_1} \propto \vecket{\mathds{1}}$, and thus we can directly add one steady state of $\hat{\mathcal{L}}^\dagger$ to the Krylov space~\footnote{Care must be taken in the presence of conserved quantities, under which the identity decomposes into sub blocks. We address this issue in~\cref{app:sec:krylov:space}.}.
\updated{
While being motivated by previous insights on the benefits of complex\hyp time Krylov\hyp spaces gained by some of us~\cite{Grundner2024,Paeckel2024}, we want to emphasize that the construction schemes introduced in this work are developed to both explore and exploit the specific structure of Lindbladian spectra.
In particular, we want to stress that the case of a real\hyp time evolution in the Lindbladian case corresponds to a sole imaginary\hyp time evolution in the Hamiltonian case.
Moreover, while in the closed system case the goal is to suppress excited states to render the time\hyp evolution tractable for tensor\hyp network methods, this angle of attack fails as soon as one is interested in the excited states of the Lindbladian.
In the latter case, the exponential decay of the low\hyp lying eigenmodes, caused by their finite negative real part, constitutes a significant challenge because the weight of these very states we try to determine vanishes exponentially fast during the time evolution.
For that purpose, and in stark contrast to the closed\hyp system approaches~\cite{Grundner2024,Paeckel2024}, the complex contours are chosen to \textit{enhance} the low\hyp lying excited states of the Lindbladian.
}
\subsection{Initial states}
\label{sec:initial:states}
\begin{figure}
    \centering
    \includegraphics[width=1\columnwidth]{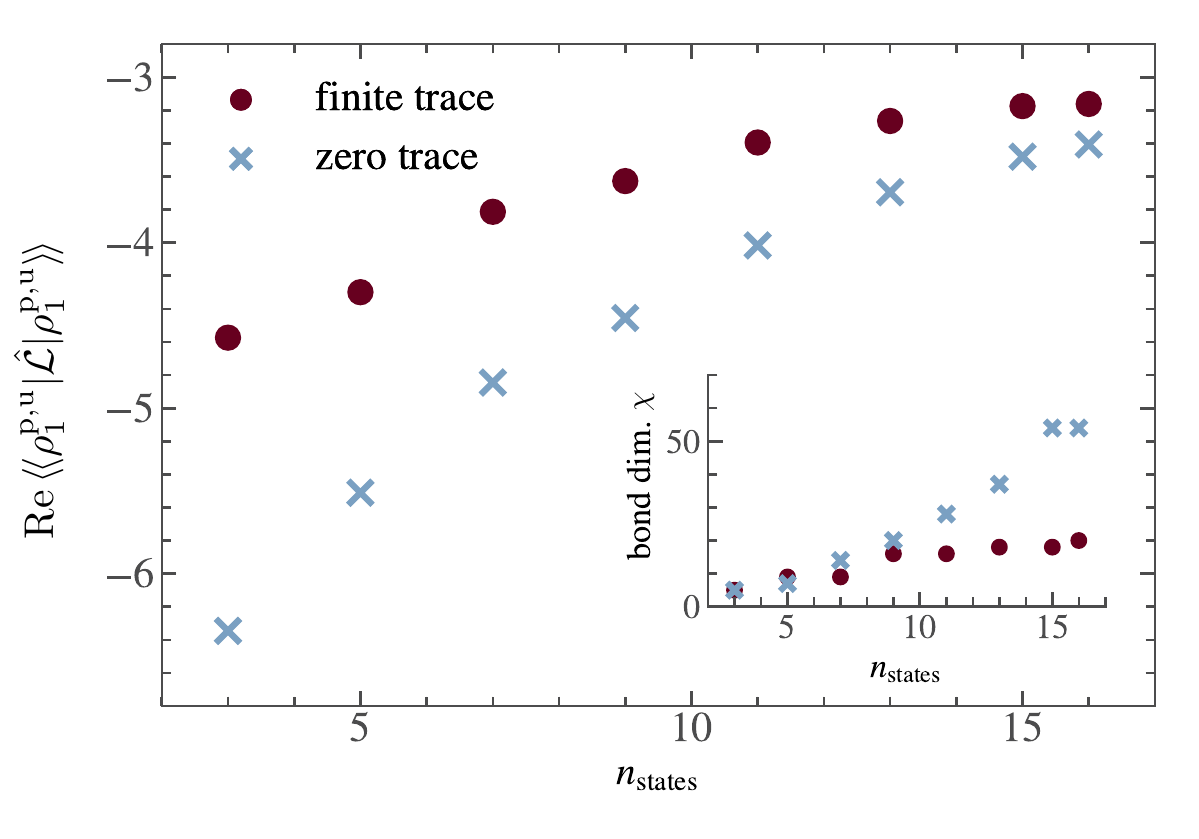}
    \caption{
        Initial states optimized via the warmup procedure. 
        We perform a search up to $n_\mathrm{states}=16$ and compute $\vecbra{\rho_1^\mathrm{p, u}}\hat{\mathcal{L}}\vecket{\rho_1^\mathrm{p, u}}$ for the optimal basis in every iteration, both for finding a physical initial state (red dots) and a traceless initial state (blue crosses).
        Inset: The corresponding maximal bond dimension of the respective states $\vecket{\rho_1^\mathrm{p, u}}$.
        All calculations were performed for a dissipative Bose Hubbard model~\cref{eq:model:hamiltonian,eq:model:jump-ops} with $L=10$ sites and $N=5$ particles, setting $J=U=\kappa=1$.
    }
    \label{fig:algorithm:warmup}
\end{figure}
The convergence rate of~\gls{CLIK-MPS}, particularly for excited states, improves significantly when initialized with optimized initial states.
Here, we construct initial states for the time evolution exhibiting a large overlap with the targeted slowly decaying modes, with the effect that the evolution times required to achieve the desired approximation qualities can be reduced.
In order to find such optimized initial guess states, we exploit the fact that the low\hyp lying eigenstates feature a small expectation value of the Lindbladian.
Moreover, we utilize different initialization procedures, specialized for either increasing the overlap with the steady state for the case of real\hyp time evolutions, or increasing the overlap with the low\hyp lying excited states for evolving along the tilted contour.
The warmup procedures are based on iteratively constructing a subspace from a set of $N_s$ pure random product states $S = \left\{ \ket{\psi_n} \right\}_n$.
We note that from any pure product state given in a~\gls{TN} representation, we can cheaply construct either traceful (physical), vectorized density operators via forming $\vecket{\sigma^\mathrm{p}_n} = \ket{\psi_n} \otimes \ket{\psi_n}$, or traceless (unphysical), vectorized density operators via forming $\vecket{\sigma^\mathrm{u}_n} = \ket{\psi_n} \otimes \ket{\varphi_n}$, where $\braket{\psi_n|\varphi_n} = 0$.
Here, the set of unphysical states is used to construct a subspace, which is in the orthogonal complement of the steady\hyp state manifold, and hence can be used to target excited states.
In principle, these vectorized states can already be used to expand the Lindbladian in the corresponding subspace by computing their matrix elements $\tilde L^\mathrm{eff}_{ij} = \vecbra{\sigma^\mathrm{u,p}_i} \hat{\mathcal L} \vecket{\sigma^\mathrm{u,p}_j}$, but the set of basis states can be improved by employing the typical dynamics generated by the Hamiltonian.
For that purpose, we introduce hoppings $\hat b^\dagger_j\hat b^\nodagger_{j+1}$ acting on a set of randomly chosen sites, which we apply to the elements of the set of product states.
Expanding $\hat{\mathcal L}$ in the obtained basis, we then solve the projected eigenvalue problem, and order the eigenvalues $\lambda_j$ increasing according to $\vert \mathrm{Re}\, \lambda_j\vert$.
After repeating this step a fixed number of times (typically $L$, the number of sites), we select the basis that yields the lowest eigenvalue $\lvert \mathrm{Re} \,\lambda_1 \rvert$.
Now we add $M$ new basis states to this optimal basis, and repeat the previous step, updating the optimal basis and the set of eigenvectors $\vecket{\rho^\mathrm{p, u}_j}$. This procedure increases the number of basis states by $M$ in each iteration (we chose $M=2$), while lowering the modulus of the lowest eigenvalue $\vert \mathrm{Re} \, \lambda_1\vert$ in the respective basis. We proceed until either a maximal Krylov dimension $n_\mathrm{states}$ is reached or the lowest eigenvector attains a certain target bond dimension. 
This defines an iterative, variational procedure with respect to the expectation value $\vecbra{\rho^\mathrm{p,u}_1} \hat{\mathcal L} \vecket{\rho^\mathrm{p,u}_1}$, which is systematically optimized towards the steady state (p case) or the slowest decaying mode (u case).
In the end, $\vecket{\rho^\mathrm{p,u}_1}$ is chosen as the optimal initial state.
In~\cref{fig:algorithm:warmup} we show the evolution of the Lindbladian expectation value for the iteratively obtained physical initial guess states $\vecket{\rho^\mathrm p_1}$ (dots) and unphysical initial guess states $\vecket{\rho^\mathrm u_1}$ (crosses) for a dissipative, interacting bosonic system with $L=10$ sites and $5$ particles.
Note the systematic improvement of the expectation value, while the bond dimensions remain comparably small (inset).
For the time evolutions, we then use $\vecket{\rho^\mathrm p_1}$ as an initial guess for the real\hyp time evolutions and $\vecket{\rho^\mathrm u_1}$ for the time evolutions following the tilted contour.
This way, we significantly enhance the Krylov subspaces, making sure that $\mathcal K_0(\rho^\mathrm p_1)$ has a large subspace overlap with the (physical) steady state, while $\mathcal K_\alpha(\rho^\mathrm u_1)$ has a large subspace overlap with the (unphysical) low\hyp lying excited states.
The detailed algorithms for the different warmup procedures are given in the \cref{app:sec:initial:states}, in which we also discuss slight modifications for the case of constructing optimized initial guess states to approximate the left eigenmodes of $\hat{\mathcal L}$. \updated{Note however that we show in \cref{sec:results} that the specific form of the warmup procedure has a subleading contribution to the observed accuracy of the method.}
\subsection{Efficient subspace arithmetic\label{subsec:subspace-expansion}}
A crucial component of~\gls{CLIK-MPS} is to avoid explicitly representing vectorized states and operators in the generated subspaces.
This can be achieved by computing the Gram matrix $\mathbf M_{ij} = \vecbraket{\rho_i}{\rho_j}$ for a given set of states $\left\{ \vecket{\rho_i} \right\}_i$, from which a transformation $\mathbf X$ into an orthonormal basis is obtained via
\begin{equation}
    \updated{\mathbf X =  \mathbf U \mathbf S^{-\nicefrac{1}{2}}}\;,
\end{equation}
where the matrices $\mathbf S,\mathbf U$ are obtained from diagonalizing the Gram matrix $\mathbf M = \mathbf U \mathbf S \mathbf U^\dagger$.
Inverting the square roots of the eigenvalues of $\mathbf M$ is a numerically delicate operation, in general.
However, using a recursive factorization~\cite{Stoudenmire2013}, the numerical precision of the computed eigenvalues can be improved to arbitrary precision such that eventually only the loss of orthogonality of the columns or rows of $\mathbf U$ imposes actual constraints.
To ensure numerical stability, we thus discard eigenvalues and eigenstates of $\mathbf M$, which violate the orthogonality constraint above a given threshold $\epsilon=10^{-14}$, i.e., perform explicit deflations.
Once we found $\mathbf X$, we can easily represent operators in the corresponding orthonormal eigenbasis by computing their expectation values $\mathbf O_{ij} = \vecbra{\rho_i} \hat{\mathcal O}\vecket{\rho_j}$ and evaluate $\mathbf O^\mathrm{eff} = \mathbf X^\dagger \mathbf O \mathbf X$.
Clearly, eigenstates of the Lindbladian are also directly obtainable by transforming expectation values $\mathbf L_{ij} = \vecbra{\rho_i} \hat{\mathcal L}\vecket{\rho_j}$ such that the representation of the Lindbladian in the Krylov subspace is given by
\begin{equation}
    \mathbf L^\mathrm{eff} := \mathbf X^\dagger \mathbf L \mathbf X \; .
\end{equation}
Denoting by $\mathbf R$ the matrix of eigenvectors of $\mathbf L^\mathrm{eff}$ and sorting their columns such that the $k$\hyp th eigenvector corresponds to the approximation of the $k$\hyp th eigenvalues, the Krylov space expansion of $\vecket{r_k}$ is given by
\begin{equation}
    \vecket{r_k} = \sum_{m=1}^{N_{\scriptscriptstyle{K}}} \left[\mathbf X \mathbf R\right]_{mk} \vecket{\rho_m} \; . \label{eq:state:expansion}
\end{equation}
Here, $N_{\scriptscriptstyle K}$ denotes the number of time\hyp evolved states and in particular we have $ \mathbf M \in \mathds C^{N_{\scriptscriptstyle K} \times N_{\scriptscriptstyle K}}$, while $D_{\scriptscriptstyle K}$ is the number of orthonormalized basis states, such that $\mathbf X \in \mathds C^{N_{\scriptscriptstyle K} \times D_{\scriptscriptstyle K}}$.
Finally, the dimension of the effective Lindbladian is given by $\operatorname{dim} \mathbf L^\mathrm{eff} = D_{\scriptscriptstyle K}$.
Again, we emphasize that~\cref{eq:state:expansion} should never be evaluated explicitly.
However, from this expansion any overlap and expectation value can be readily derived in such a way that only arithmetics with states and operators expressed in terms of the states $\vecket{\rho_i}$ is needed.
Several algorithmic optimizations can be exploited to further improve the computational efficiency, using algebraic properties of states and operators in the Krylov space.
In~\cref{app:sec:expectations:and:overlaps,sec:app:discrete:symmetries}, we elaborate on the specific implementations in detail and elucidate the role of Lindbladian symmetries. 
Here we note that steady states must be hermitian, yet the complex\hyp time evolution can introduce non\hyp Hermitian artifacts due to truncation and approximation errors of the time\hyp evolution scheme used, as well as from the complex time contour itself.
These non\hyp Hermitian perturbations can be easily removed by hermitizing the representation of the density operators $\vecket{\rho} \rightarrow (\vecket{\rho} + \vecket{\rho^\dagger})/2$ used to evaluate expectation values.
Again, exploiting~\cref{eq:state:expansion}, this can be directly incorporated on the level of the Krylov space representation and no arithmetics in terms of tensor\hyp network states is required.
\subsection{\updated{High level structure}}
\updated{
To summarize the whole \gls{CLIK-MPS} algorithm, we provide a high\hyp level pseudocode in \cref{alg:clik:mps}.
}
\updated{
Given a Lindbladian $\hat{\mathcal L}$, we first approximate $\alpha_\mathrm{max}$ as briefly described in \cref{subsec:krylov:space} and detailed in \cref{app:alg:find:alpha}. Now, a sensible angle for the linearly inclined contour  $\alpha \ll \alpha_\mathrm{max}$ can be chosen. 
Two initial states are generated using the initialization procedures specified in \cref{sec:initial:states} and in \cref{app:alg:iterative:initial}. 
They are a physical ($\vecket{\rho_\mathrm{p}}$) and a traceless ($\vecket{\rho_\mathrm{u}}$) one. 
We then start two time evolutions in parallel, a real time evolution where we save every step $\delta t$ of the physical, as well as a complex time evolution with step $\delta z$ of the traceless one.
After each timestep, the evolved states are appended to the Krylov space $\mathcal K_{0, \alpha}$.
Note that the overlaps and expectations needed to express the Lindbladian inside of the Krylov space can be calculated in parallel as soon as a new timestep has been performed, as detailed in \cref{app:alg:krylov:subspace}. 
For this, only a few matrix elements need to be calculated in each step, reducing the number of expectation values in each step to linear in the number of time evolved states (this is indicated by the * in \cref{alg:clik:mps}).
Tracking the spectrum of the Lindbladian expressed inside the Krylov space over time can be used to establish escape criteria for the time evolution. 
This can also be generalized to the convergence of expectations of observables in the steady state, for instance. A sensible choice of the escape criterion depends on the specific use case.
Note that we did not use any automated escape criteria to generate the results provided in \cref{sec:results} to better verify the results we got.}
\updated{
\begin{algorithm}[H]
\caption{CLIK-MPS algorithm}
\label{alg:clik:mps}
\begin{algorithmic}[1]
\Procedure{\updated{CLIK-MPS}}{\updated{$\hat{\mathcal L}$}}
    \State \updated{$\alpha_\mathrm{max}$ $\leftarrow$ from \cref{app:alg:find:alpha}}
    \State \updated{$\delta z = \mathrm e^{-\mathrm i \alpha}\delta t$ $\leftarrow$ with $\alpha \ll \alpha_\mathrm{max}$}
    \State \updated{$\vecket{\rho_\mathrm{p}}, \vecket{\rho_\mathrm{u}}$ $\leftarrow$ from \cref{app:alg:iterative:initial}}
    \For{\updated{$n < n_\mathrm{max}$}}
        \State \updated{$\vecket{\rho_\mathrm{p}(\delta t n)}$ $\leftarrow$ do and save timestep}
        \State \updated{$\vecket{\rho_\mathrm{u}(\delta z n)}$ $\leftarrow$ do and save timestep}
        \State \updated{Append states to Krylov space $\mathcal K_{0, \alpha}$}
        \State \updated{$\{\lambda_i, \mathbf r^i\}_i $ $\leftarrow$ do \cref{app:alg:krylov:subspace} with $\mathcal K_{0, \alpha}$ (*)}
        \State \updated{Break if $\mathcal K_{0, \alpha}$ is converged}
    \EndFor
\EndProcedure
\end{algorithmic}
\end{algorithm}
}
%

%
\section{Results\label{sec:results}}
\label{sec:results}
We apply~\gls{CLIK-MPS} to study a driven system of interacting bosons on a \gls{1D} lattice.
The Hamiltonian is given by
\begin{equation}
    \hat H = -J \sum_{j=1}^{L-1} \left( \hat{b}^\dagger_{j+1} \hat{b}^\nodagger_{j} + \mathrm{h.c.} \right) + \frac{U}{2} \sum_{j=1}^{L} \hat{b}^{\dagger 2}_j \hat{b}^{\nodagger 2}_j \;,
    \label{eq:model:hamiltonian}
\end{equation}
where, $U$  and $J$ are the onsite interaction strength and nearest\hyp neighbor hopping amplitude, respectively, and $\hat{b}^\nodagger_j$ ($\hat{b}^\dagger_j$) annihilates (creates) a boson on site $j$.
The dissipative dynamics are generated by the jump operators
\begin{equation}
    \hat{L}_j = \sqrt\kappa(\hat{b}^\dagger_{j+1} + \hat{b}^\dagger_{j})(\hat{b}^\nodagger_{j+1} - \hat{b}^\nodagger_{j}) \;,
    \label{eq:model:jump-ops}
\end{equation}
where $\kappa$ is the dissipation strength.
The resulting Lindbladian has been shown recently to exhibit a local~\gls{BEC} with very slowly decaying spatial correlations~\cite{Kraus2008, Diehl2008,westhoff2025fast}, providing an ideal testing platform as a strongly correlated, driven system.
Physically, the system\hyp environment couplings in this model exhibit an interesting feature: They are non\hyp local in real space, yet coupled to a Markovian environment.
This situation can be made more transparent by transforming the model~\cref{eq:model:hamiltonian}, as well as the \updated{jump operators}~\cref{eq:model:jump-ops} to momentum space.
Then, the system realizes a local coupling for the bosonic, single\hyp particle momentum\hyp space modes $q$, which now are individually coupled to a structured environment, i.e., the couplings become momentum dependent with a spectral density depending on the momentum $k$ given by
\begin{equation}
    \big\vert g_q(k)\big\vert ^2 = 16 \cos\Big( \frac{k-q}{2}\Big)^2 \sin\Big( \frac k 2\Big)^2 \; .
\end{equation}
Such structured spectral densities are a typical feature of non\hyp Markovian environments and in general occur when incorporating Markovian embeddings to describe non\hyp Markovian environments~\cite{Moroder2022,Xie2024}.
We thus study a problem providing a numerical challenge comparable to what can be expected for Markovian embeddings of non\hyp Markovian setups, while still allowing us to assess the approximation quality of physical observables by comparing to analytically and numerically known results~\cite{Kraus2008, Diehl2008,westhoff2025fast}.
Moreover, general non\hyp Markovian systems can readily be described by replacing the Lindblad generator $\hat{\mathcal{L}}$ with the \gls{HEOM} generator $\hat{\mathcal{L}}_\mathrm{HEOM}$~\cite{Tanimura2020, Debecker2024Spectral}, which features the same spectral properties.
\subsection{Comparison with~\acrshort{ED}}
\begin{figure}
    \centering
    \subfloat[\label{subfig:mps:ed:comparison:eigenvalues}]{
        \includegraphics[width=0.97\columnwidth]{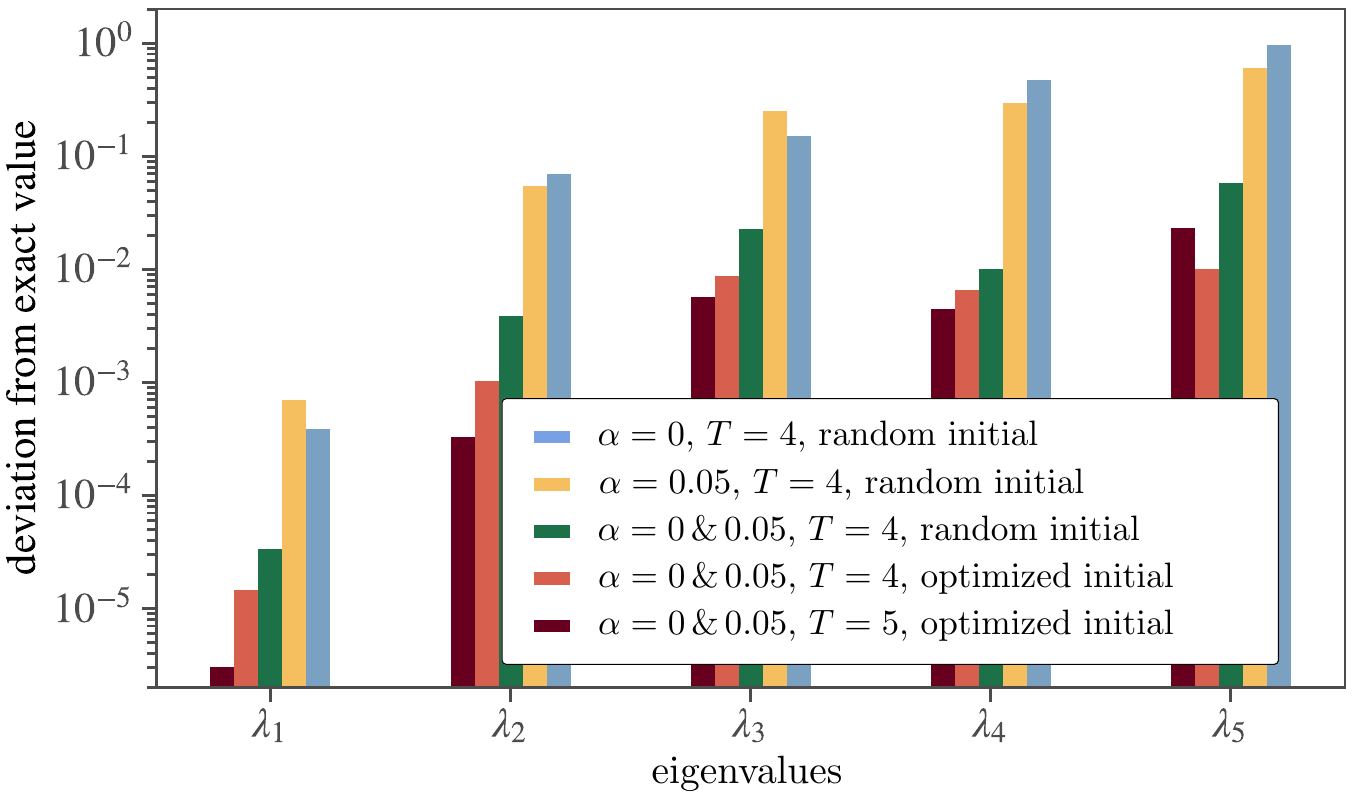}
    }
    \vspace{-0.4cm} 
    \subfloat[\label{subfig:mps:ed:comparison:reigenvectors}]{
        \includegraphics[width=0.473\columnwidth]{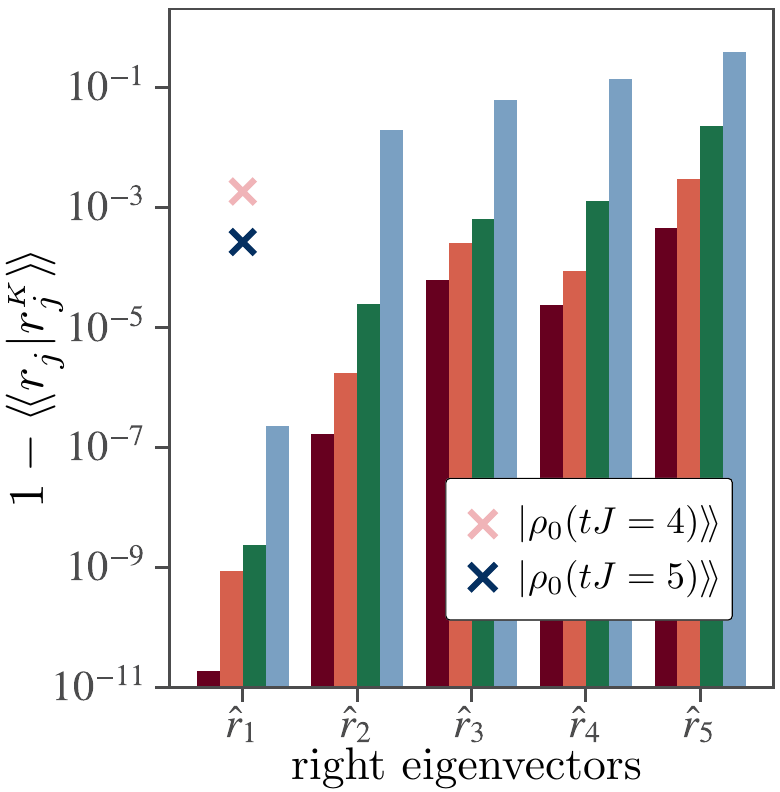}
    }
    \hspace{-0.2cm} 
    \subfloat[\label{subfig:mps:ed:comparison:leigenvectors}]{
        \includegraphics[width=0.473\columnwidth]{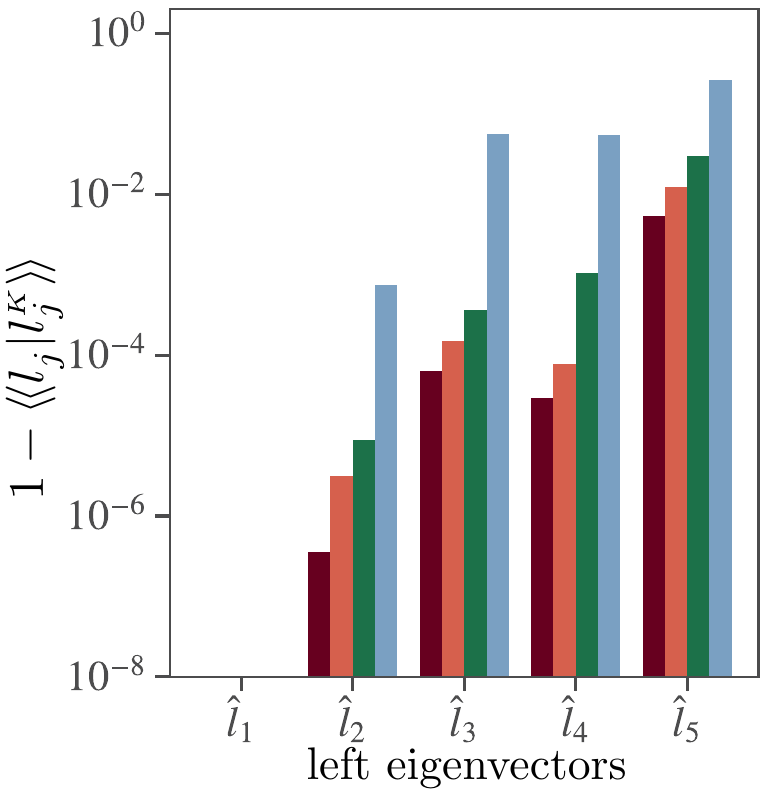}
    }
    \caption{
        Accuracy of~\gls{CLIK-MPS} compared against~\gls{ED} and real\hyp time evolution methods~\cite{Minganti2022}.
        In \cref{subfig:mps:ed:comparison:eigenvalues}, we compare the lowest Lindbladian eigenvalues \updated{from different Krylov based methods} with \gls{ED} reference data.
        \updated{We consider time evolution with $\hat{\mathcal{L}}$ for a real\hyp time evolution method (blue) as well as a complex time evolution only (yellow) compared to \gls{CLIK-MPS} with random (green) and optimized (orange) initial states. We plot the deviation of the approximated eigenvalues to the exact data.}
        In \cref{subfig:mps:ed:comparison:reigenvectors} we compare the deviations of the Linbladian right eigenmodes for real and complex time evolutions with $\hat{\mathcal{L}}$ for different final times.
        The deviations of the time\hyp evolved state at the final evolution time from the exact steady states are also indicated (\protect\blackcross).
        In \cref{subfig:mps:ed:comparison:leigenvectors}, the same analysis is shown for the left eigenmodes obtained through time evolution with $\hat{\mathcal L}^\dagger$.
        We considered $L=5$ lattice sites, $N=4$ particles, set $J=U=\kappa=1$, employed a timestep $\mathrm d t=0.05$ during time evolution and used every second time\hyp evolved state for the Krylov space. The maximal evolution time $T$ and the complex angles $\alpha$ are specified in the plot. 
        \updated{For all data relying on random initial states we average over 50 such initializations.}
        Details on the choice of initial states are given in the main text. For the data with $\hat{\mathcal L}$, the optimized initial states (orange and red) had expectation values $-1.84$ for $\alpha=0$ and $-3.32$ for $\alpha=0.05$.
    }
    \label{fig:mps:ed:comparison}
\end{figure}
We begin analyzing the convergence properties of~\gls{CLIK-MPS} by simulating small realizations of~\cref{eq:model:hamiltonian,eq:model:jump-ops}, which can be compared to~\gls{ED} results.
To demonstrate the improvements, we compare to results obtained using a Krylov space generated from a real\hyp time evolution of a random initial state, which was suggested very recently and, to the best of our knowledge, provides the most generic and efficient approach to study Lindbladian spectra so far~\cite{Minganti2022}.
In order to achieve a fair comparison to this real\hyp time approach, which we will also refer to as the $\alpha=0$ method in the following, we use the same maximum evolution times when comparing $\alpha=0$ results to~\gls{CLIK-MPS} results. 
\updated{Additionally, to analyze the impact of the different components of \gls{CLIK-MPS}, we include a Krylov space with both a real and complex time evolution, but without the warmup procedure, which we will refer to as \textit{random initial CLIK-MPS}, as well as a Krylov space generated from a complex time evolution alone.}
In~\cref{subfig:mps:ed:comparison:eigenvalues} \updated{we show the first Krylov low\hyp lying eigenvalues for different methods. In order to assess the performance of the methods, we plot the difference of the Krylov eigenvalues to the exact data from \gls{ED}.
The steady state ($\lambda_1 = 0$) is reproduced seemingly well using either the $\alpha=0$ method (blue), a single complex time evolution (yellow), or~\gls{CLIK-MPS} (orange), as well as the random initial CLIK-MPS (green).
However, already for the first excited state, the results from single time evolutions have a deviation in the order of $10^{-1}$, which is quite large compared to the eigenvalue $\lambda_1 = -0.84 + 0.81 \, \mathrm i$, and gets even more severe when trying to evaluate higher excited states.
On the other hand~\gls{CLIK-MPS} exhibits a robust and precise approximation of the first four low\hyp lying excited states. Similarly, the random initial \gls{CLIK-MPS} shows a comparable approximation quality like \gls{CLIK-MPS}, indicating that the main benefit originates from combining real and complex time evolution and not the choice of the warmup procedure.
}
Note that the chosen positive angle $\alpha=0.05$ enhances the precision of the eigenvalues with \updated{positive} imaginary part and decreases the accuracy of those with negative imaginary part.
However, the lower half\hyp plane eigenvalues can be obtained simply by hermitian conjugation of the eigenvalues located in the upper half\hyp plane.
Next, we compute the deviation of the Krylov Lindbladian right and left eigenvectors $\vecket{r^\text{\tiny{K}}_j}$, $\vecket{l^\text{\tiny{K}}_j}$ from the \gls{ED} reference data in \cref{subfig:mps:ed:comparison:reigenvectors} and \cref{subfig:mps:ed:comparison:leigenvectors}, respectively.
The computation of these overlaps is detailed in~\cref{app:sec:ed:comparison}.
Here, the impact of the additional complex\hyp time evolution becomes evident when comparing with the $\alpha=0$ curves: The accuracy is improved by more than three orders of magnitude! \updated{The improvements are similar when considering the random initial CLIK-MPS (green).}
Note that here we also include a comparison against another, even simpler approach to find the steady state, namely approximating it by a real\hyp time evolution, only (crosses).
This ansatz exhibits the poorest convergence when evolving up to the same maximal evolution time as the one used for the Krylov\hyp space\hyp based methods, with a loss in precision of more than six orders of magnitude.
Moreover, increasing the duration of the time evolution systematically enhances the precision not only of the steady state $\vecket{r^\text{\tiny{K}}_1}$, but of all right eigenvectors.
For the Krylov left eigenvectors $\vecket{l^{\scriptscriptstyle K}_j}$ in \cref{subfig:mps:ed:comparison:leigenvectors}, note that the three slowest decaying left eigenmodes are captured with a precision of at least $10^{-4}$ with the slowest decaying one even at $10^{-6}$.
\subsection{Large\hyp scale systems beyond~\acrshort{ED}}
We now turn to an in\hyp depth analysis of the phase diagram of~\cref{eq:model:hamiltonian} driven by the jump operators~\cref{eq:model:jump-ops}.
From previous works~\cite{Kraus2008, Diehl2008,westhoff2025fast}, the model is known to exhibit a steady state realizing a so\hyp called \textit{local~\gls{BEC}} in the limit of weak interaction strength $U/J \ll 1$, or strong dissipation $\kappa/J \gg 1$.
Here, \textit{local} refers to the observation that in these limits, the correlation length increases rapidly such that very large system sizes are required to discriminate the exponential decay of particle\hyp correlations.
In the following, we use~\gls{CLIK-MPS} to study the model in a large parameter regime, putting particular emphasis on this crossover where the correlation length competes with the system size.
Furthermore, we show how~\gls{CLIK-MPS} can be used to evaluate the so\hyp called Mpemba speedup, which provides a tool to speed up the experimental preparation time of the steady state exponentially, rendering the practical realization of~\cref{eq:model:hamiltonian,eq:model:jump-ops} a promising candidate for a platform to speed up the preparation of~\glspl{BEC} exponentially.
Note that both types of analysis require the computation not only of the steady state but also the Lindbladian gap \updated{as well as higher eigenvalues}.
Therefore, these investigations also illustrate the physical relevance of the ability to evaluate both the steady state, as well as the low\hyp lying excited states with high precision.
Finally, we also performed extensive and systematic convergence analysis for the data provided, which we discuss in depth in~\cref{app:sec:convergence:large:systems}, and prove an exact error bound on the steady state, as well \updated{as Krylov error bounds for the convergence of eigenvectors}.
\subsubsection{Correlation length and local~\gls{BEC}}
\begin{figure}[t]
    \centering
    \includegraphics[width=0.9\columnwidth]{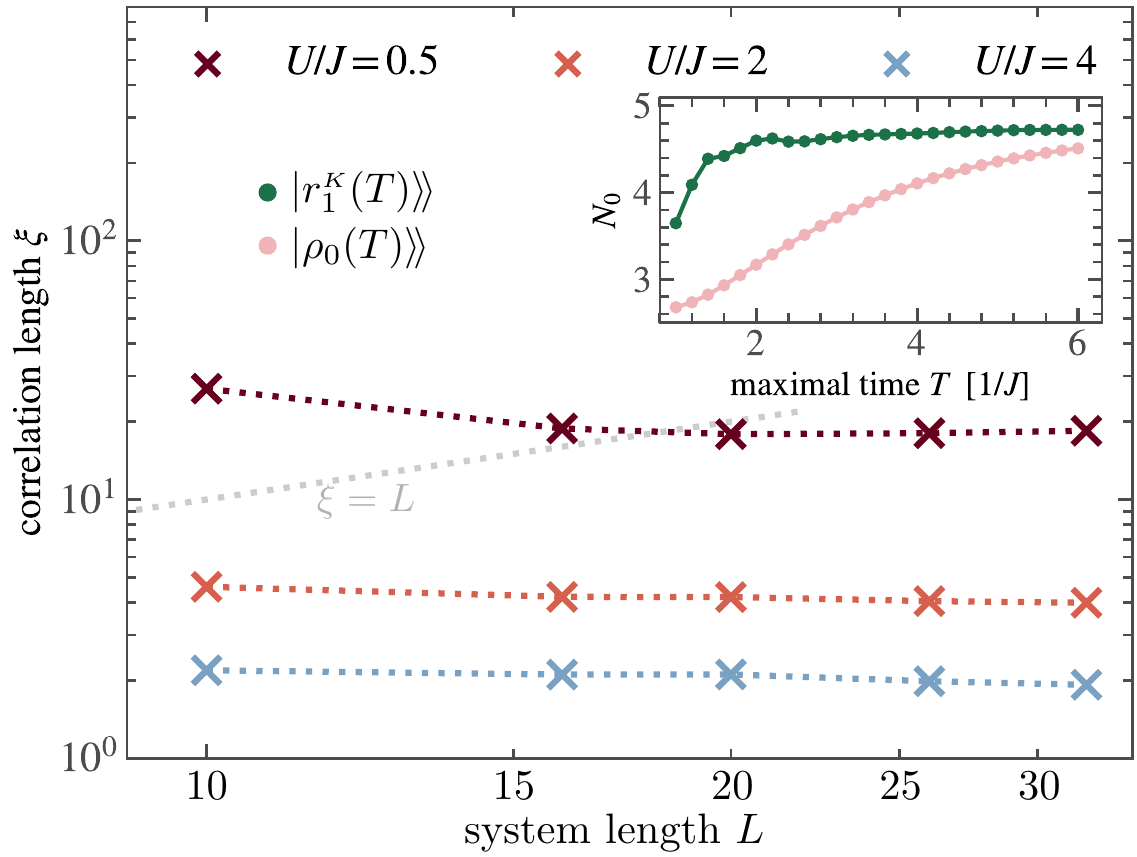}
    \caption{
        Correlation lengths $\xi$ (\cref{eq:correlation-length}) of the dissipative Bose\hyp Hubbard steady state calculated within \gls{CLIK-MPS} for different lengths and onsite interactions $U$.
        $\xi$ is almost independent of the length of the chain and shrinks upon increasing $U$. Note that if $\xi>L$, the system is in a lattice analog of a \gls{BEC}. 
        Inset: Leading eigenvalue $N_0$ of the correlation matrix $C(i-j)$ depending on the evolved time $T$.
        We compare the real\hyp time evolved state at time $T$ (pink) to the \gls{CLIK-MPS} steady state approximation (green).
        We show data for a system with $L=10$ sites at half filling, with $U=0$, $\kappa/J=2$.
        \newline At $L=32$ we used $T=32$ and $\mathrm \delta t=0.4$, at $L=26, \; T=22$ and $ \mathrm \delta t=0.4$, at $L=20, \;T=18$ and $ \mathrm \delta t=0.2$, at $L=16 , \; T=10 $ and $ \delta t=0.1$, while at $L=10$ we used $T=6, \, \mathrm \delta t=0.1$.
        For $L=32$, $L=26$, $L=20$, and $L=16$ sites we used the complex angle $\alpha=0.02$, while for $L=10$ we employed $\alpha=0.05$.
    }
    \label{fig:steady-state:correlation-lengths}
\end{figure}
We begin by studying the crossover behavior at repulsive interaction strengths $U$ in the steady state at half filling $N=L/2$.
We compute the correlation length $\xi$, which can be inferred from the two\hyp point correlator $C(i-j)=\langle \hat b_i^\dagger \hat b_{j}^\nodagger\rangle_\mathrm{ss}$ using a fitting function that captures an exponential decay in the long\hyp distance limit
\begin{equation}
    C(i-j) = \frac{c_1}{\vert i-j\vert^{c_2}}\mathrm e^{-|i-j|/\xi} +c_3 \,.
    \label{eq:correlation-length}
\end{equation}
We fix $i=L/4$ to counteract boundary effects and fit the tail of $C$.
In~\cref{fig:steady-state:correlation-lengths} we show the extracted correlation lengths as a function of the number of lattice sites, varying the Hubbard interaction strengths, for a fixed dissipation $\kappa/J=2$.
The dependence of the correlation length on the system size $L$ indicates the formation of a local \gls{BEC} for small Hubbard repulsion strengths $U$.
However, it is important to note that fitting the exponential decay can be misleading if the system size is too small.
To illustrate this point, we indicate the case $\xi = L$ by the gray line.
Any correlation length fitted above this line, i.e., $\xi>L$, should not be trusted because the system is too small to recognize the finite correlation length.
On the other hand, this is exactly the region where the local~\gls{BEC} appears: The system behaves as a~\gls{BEC} on all accessible length scales.
To illustrate the superior convergence of~\gls{CLIK-MPS}, in the inset of~\cref{fig:steady-state:correlation-lengths} we compare the convergence of the leading eigenvalue of the correlator $C(i-j)$ to the values obtained using time evolved states, as a function of the maximum evolution time.
This eigenvalue characterizes the condensate fraction of the bosonic particles and, hence, serves as an excellent proxy to check numerical convergence.
Computing $C(i-j)$ from the right eigenstates obtained using~\gls{CLIK-MPS} (green symbols), we can clearly observe a rapid saturation already at times $T=4$.
In contrast, the real\hyp time evolution method converges significantly slower (pink symbols) and considerably longer simulation times are required.
\subsubsection{Dissipative gap and critical point at $U=0$}
\begin{figure}
    \centering
    \includegraphics[width=0.95\columnwidth]{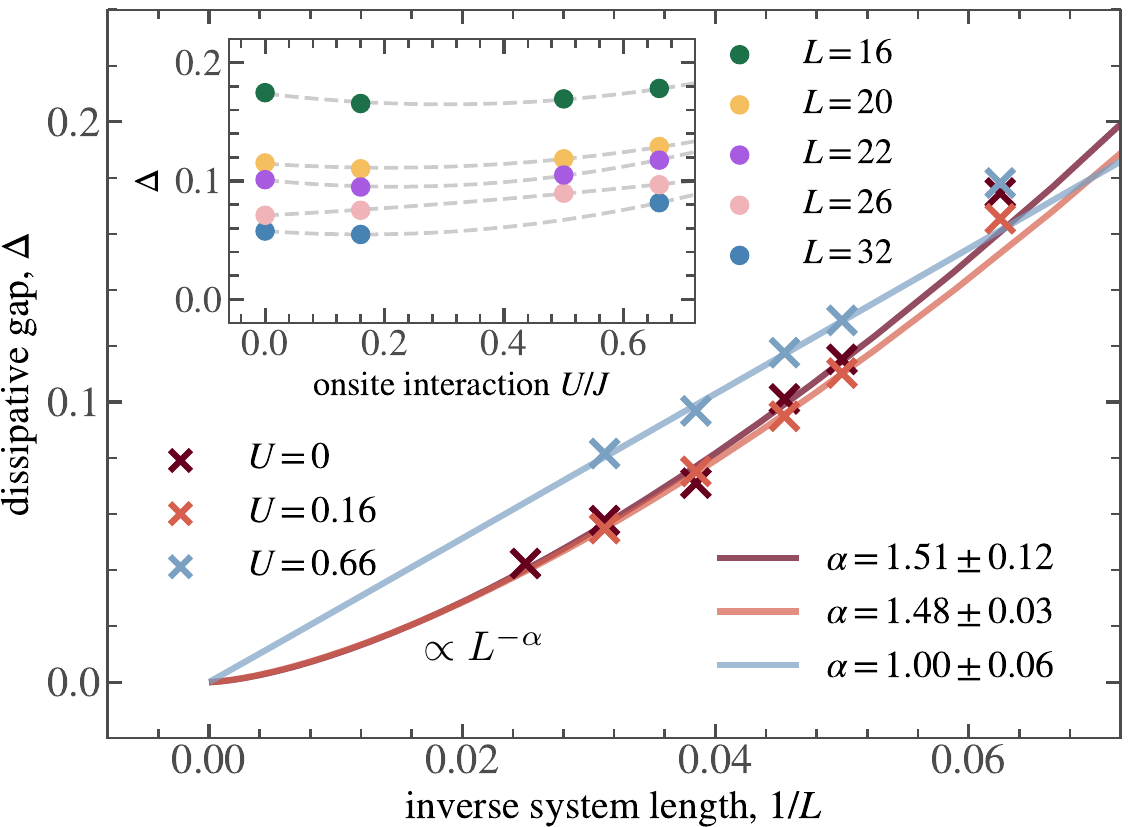}
    \caption{
        \label{fig:spectrum:dissipative-gap}
        Scaling of the dissipative gap at half filling.
        We show the dissipative gap $\Delta=-\mathrm{Re}\,\lambda_2$ for three onsite interactions at $\kappa/J=2$ depending on the inverse system size $1/L$. We perform a fit of the data according to \cref{eq:gap:ansatz}, where we exclude the $L=16$ data points. 
        We find a power\hyp law dependence $\Delta \propto L^{-\alpha}$ with the scaling exponent $\alpha$ increasing from $\alpha=1$ towards $\alpha \approx 3/2$ when $U/J\rightarrow 0$.
        Inset: Dissipative gap $\Delta$ for several system sizes and onsite interactions. At these small onsite interactions, the gap depends quadratically on $U$.
        For $L=16$ and $20$, we use $\delta t=0.2$ and $T=18, \, 28$, respectively, while for $L=22$ and $26$ we have $\delta t=0.4$ and $T=40$ and $50$, respectively. 
        For $L=32$ \updated{and $L=40$}, $\delta t=0.6$ and $T$ was chosen $U$\hyp dependent to ensure convergence (see \cref{app:sec:convergence:large:systems}, for \updated{$L=40$, $T=120$}).
    }
\end{figure}
It is well\hyp known that~\cref{eq:model:hamiltonian,eq:model:jump-ops} exhibits a critical point at $U=0$~\cite{Kraus2008, Diehl2008,westhoff2025fast}, where a true~\gls{BEC} is stabilized.
This result can already by inferred from~\cref{fig:steady-state:correlation-lengths}, where we observe an increasing correlation length with decreasing Hubbard interaction.
Here, we investigate this crossover more systematically by studying the system\hyp size and interaction dependency of the dissipative gap $\Delta = -\operatorname{Re} \lambda_2$.
%
%
Since the limit $U/\kappa\rightarrow 0$ can be treated analytically~\cite{westhoff2025fast}, the situation is more controlled in our case, and we leave the demonstration that~\gls{CLIK-MPS} can be used to reliably identify~\glspl{DPT} to future works.
In~\Cref{fig:spectrum:dissipative-gap}, we show the dissipative gap for various repulsive Hubbard interaction strengths as a function of the system size, keeping the particle density constant: $N/L = 0.5$. \updated{We consider systems of up to $40$ bosonic sites with local dimension $21$, resulting in Hilbert spaces of dimension $\mathrm{dim}\, \mathcal H \approx 2^{176}$.}
For the gap, it is known that $\Delta\rightarrow 0$ for $L\rightarrow\infty$~\cite{westhoff2025fast} and the asymptotic scaling relation is of the form~\cite{vznidarivc2015relaxation}
\begin{equation}
     \Delta(L) = \frac{a}{L^\alpha}\; .     \label{eq:gap:ansatz}
\end{equation}
%
We hence use this ansatz to infer the scaling of the gap in the infinite system limit.
In~\cref{fig:spectrum:dissipative-gap}, we show gaps evaluated for various values of the Hubbard interaction and system sizes, as well as the resulting finite\hyp size extrapolations.
%

Note that for the fits we had to exclude the smallest systems with $L=16$ sites in order to reliably identify the gap closing, which is perfectly consistent with the observed correlation lengths in~\cref{fig:steady-state:correlation-lengths}.
%
\updated{Interestingly, we observe that for intermediate interactions ($U = 0.66$), the gap scales linearly with inverse system size, $\Delta \sim L^{-1}$, corresponding to a relaxation time $\tau = 1/\Delta \sim L$. 
In contrast, for weak or vanishing interactions ($U = 0$ and $U = 0.16$), the gap closes more rapidly with system size, following $\Delta \sim L^{-3/2}$ within numerical accuracy. 
This corresponds to a relaxation time scaling as $\tau \sim L^{3/2}$ and indicates superdiffusive relaxation dynamics. 
The appearance of the exponent $3/2$ is particularly noteworthy, as it coincides with the dynamical exponent of \gls{KPZ} scaling~\cite{Hart2015}, revealing \gls{KPZ}-type superdiffusive relaxation in a dissipative many-body system.
Note that although we considered large system sizes, there is the possibility of a change of convergence speed due to an eigenvalue crossing at higher system sizes.}
%

%
The observation that for a local~\gls{BEC} the relaxation time $\tau = 1/\Delta$ increases faster than linear with the system size seems to complicate the practical realization due to a rapidly growing time scale.
However, this problem can be overcome by choosing special initial states.
While it has been known for some time that the relaxation speed of driven systems can exhibit anomalous behavior for specific initial states~\cite{Mori2020}, in general, identifying such states is highly non\hyp trivial.
In the next paragraph, we demonstrate how the capability to access several low\hyp lying Lindbladian eigenstates allows us to address this issue, and we show that there is a class of experimentally simple\hyp to\hyp realize states for which relaxation times can be sped up exponentially.
\subsubsection{Anomalous thermalization and the Mpemba effect}
\begin{figure}
    \centering
    \includegraphics[width=0.95\columnwidth]{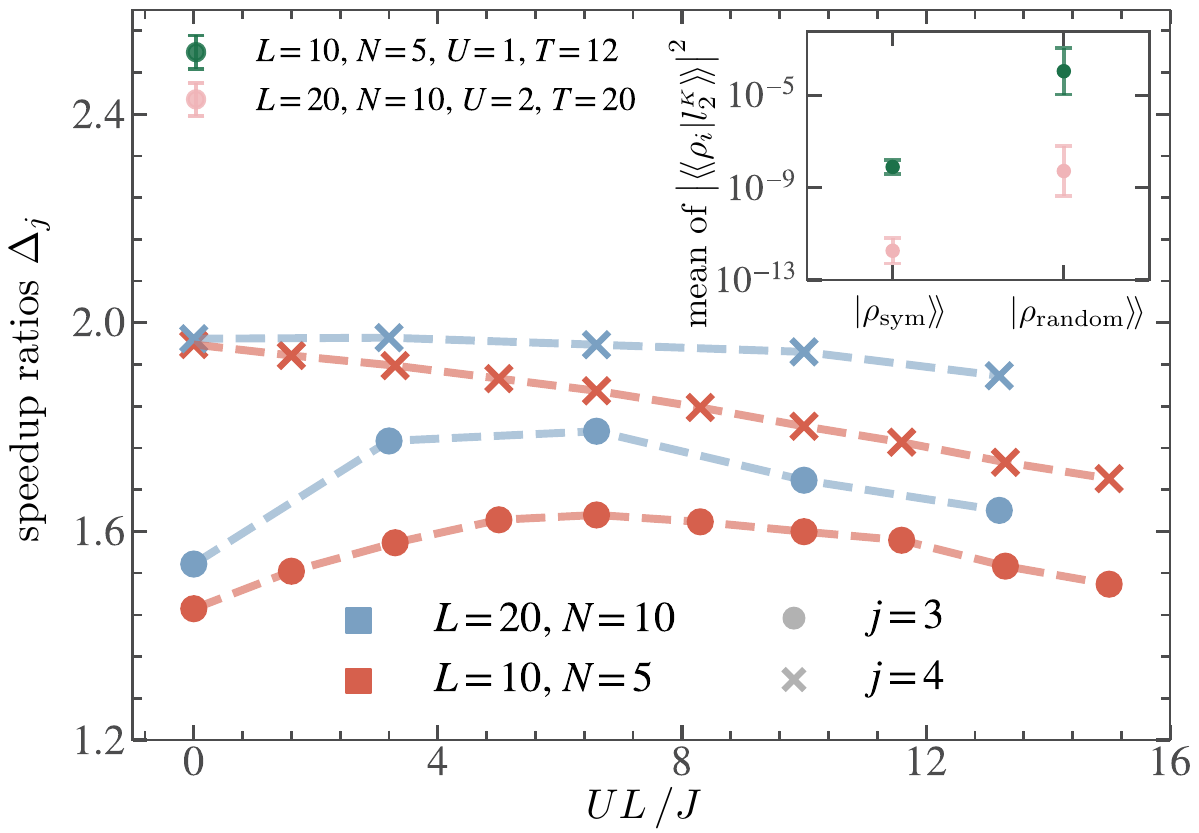}
    \caption{\label{fig:spec-and-levecs}
        The Mpemba\hyp speedup ratios $\Delta_j =\mathrm{Re}\lambda_j/\mathrm{Re}\lambda_2$ (see \cref{eq:relative:gap}) as a function of the onsite interaction $U$ for two lengths of the chain. 
        Inset: Mean overlaps of different product states with the obtained slowest decaying left eigenmode $\vecket{l^{\scriptscriptstyle K}_2}$. The mean was calculated with 50 randomly sampled symmetric states ($\vecket{\rho_\mathrm{sym}}$) and random product states ($\vecket{\rho_\mathrm{random}}$), respectively.
        For symmetric states, the overlap is 4 orders of magnitude smaller, indicating the emergence of the~\acrshort{QME}. The error bars show that this behavior is systematic.
        For all calculations, we use a local dimension $d=N+1$, while we set the maximal bond dimension $\chi=2400$ during time evolution. For data with $L=20$ and $L=16$, we use a Krylov timestep of $\mathrm \delta t=0.2$ and complex angle $\alpha=0.02$, while for $L=10$ we chose $\mathrm \delta t=0.1$ and $\alpha=0.05$. The maximal time $T$ is set to $T=12$ for $L=10$, while it depends on $U$ for $L=20$ and was chosen such that the results are sufficiently converged (see \cref{app:sec:convergence:large:systems}).
    }
    \label{fig:gap:and:speedups}
\end{figure}
We now demonstrate how the capability to access spectral regions beyond the steady and first excited state allows us to determine practically relevant, physical quantities.
For that purpose, we study the emergence of anomalous equilibration processes at the example of the~\gls{QME} \cite{Lu2017, Carollo2021, Ares2023Nat, Moroder2024, Beato2025, Summer2026} by computing the relative speedup
\begin{equation}
    \Delta_j= \frac{\mathrm{Re}\, \lambda_j}{\mathrm{Re} \,\lambda_2} \,.
    \label{eq:relative:gap}
\end{equation}
This quantity determines the magnitude of the exponential speed up in relaxation time towards the steady state for special initial states $\vecket{\rho_0}$ compared to random ones, assuming that $\vecket{\rho_0}$ exhibits zero overlap with the slowest\hyp decaying mode $\vecbraket{\rho_0}{l_2} = 0$.
From an experimental point of view, such an exponential speed-up in the preparation time constitutes an important improvement, since long preparation times are the main limiting factor for high fidelities of the desired target state~\cite{Schlosshauer2019}.
In~\cite{westhoff2025fast}, it was established using~\gls{ED} methods that there are easy\hyp to\hyp prepare states $\vecket{\rho_\mathrm{sym}}$, which can be constructing as product states that are invariant under mirroring about the center of the lattice, and feature exponentially faster relaxation (for instance $\vecket{1,1,\dots,1,1}$ and $\vecket{1,2,\dots,2,1}$).
In~\cref{fig:spec-and-levecs}, we show the dependence of $\Delta_j$ on the repulsive on\hyp site interaction strength $U$ for various system sizes at half filling.
The required eigenvalues are calculated using \gls{CLIK-MPS}.
While $\Delta_4$ decays monotonically, yet slowly, for $\Delta_3$ we observe a pronounced maximum near $UL/J \approx 6.5$ with a speed up of $\sim 1.5$ orders of magnitude.
Note that this value increases with the system size.
In order to confirm that indeed there is a~\gls{QME}, we also computed the overlap of $\vecket{\rho_\mathrm{sym}}$ with the slowest decaying mode $\vecket{l_2}$.
In the inset of~\cref{fig:spec-and-levecs}, we show this overlap (violet points) and compare it to the overlap of random initial product states $\vecket{\rho_\mathrm{random}}$, each averaged over $50$ realizations.
For the symmetric initial state, this overlap is $4-6$ orders of magnitude smaller than for the random initial states.
This extends the previous findings~\cite{westhoff2025fast}, in which the~\gls{QME} could be verified for much smaller systems only.
The fact that for finite interaction strengths the relaxation of a many\hyp body system towards a steady state with correlation lengths beyond the lattice size can be sped up exponentially is intriguing.
Apparently, for symmetric initial configurations, the Hubbard repulsion enhances incoherent scattering in the scope of the dynamics, yet incoherences are strongly suppressed once the system has thermalized.
This observation highlights the important role of the complex interplay between interactions and transport properties in driven systems, and how analyzing spectral regions of the Lindbladian beyond the steady state can help to disentangle their intricate relationship.
\section{Conclusion}
\label{sec:conclusion}
We introduced \gls{CLIK-MPS}, a \acrlong{TN}\hyp based framework that not only enables the efficient computation of the steady state and the dissipative gap, but also to explore the low\hyp lying Lindbladian eigenvalues and eigenmodes with high accuracy.
The framework is built on recent developments of complex\hyp time Krylov space methods and expands these ideas by specific complex contours and efficient construction schemes for vectorized density matrices and superoperators in high\hyp dimensional Krylov subspaces.
The resulting time\hyp evolution schemes are combined with specific initialization methods to obtain optimized initial states, such that the subsequent dynamics yield Krylov\hyp spaces capturing the low\hyp lying Lindbladian eigenstates for Hilbert space dimensions, which have been far beyond reach for previous approaches.
\updated{We note that recently a variant of the ESPRIT algorithm~\cite{Paulraj1986,Weilin2020} was used to calculate Lindbladian eigenvalues ~\cite{Park2024, Huang2026}, yet for significantly smaller system sizes than those analyzed here.}
In order to ascertain its capabilities, we used~\gls{CLIK-MPS} to investigate a driven Bose\hyp Hubbard model featuring a non\hyp trivial, structured spectral density between the bosonic momentum\hyp space single\hyp particle modes, and the environment.
This type of system\hyp environment coupling is reminiscent of non\hyp Markovian environments, described via Markovian embeddings and, thereby, provides an excellent testing ground for the computational challenges occurring for non\hyp Markovian problems.
At the example of small systems amenable to~\gls{ED}, we benchmarked the performance of~\gls{CLIK-MPS} against a recently proposed Krylov\hyp space approach~\cite{Minganti2022}, finding an improvement in the accuracy of nearly four orders of magnitude, when evaluating excited Lindbladian left\hyp and right\hyp eigenmodes.
Exploiting this striking accuracy and the high computational efficiency, we in\hyp depth investigated the emergence of a local~~\gls{BEC}, which is characterized by an increasing correlation length for small, repulsive Hubbard interactions culminating to a critical point at $U/J=0$ where a true~\gls{BEC} is stabilized.
Examining this crossover behavior is numerically challenging due to the competition between the growing correlation length $\xi$ and the closing of the dissipative gap $\Delta$, which vanishes when increasing the system size.
Using large-scale numerics with up to \updated{$L=40$} lattice sites and $N=L/2$ bosons, \updated{leading to Hilbert space dimensions of $\mathrm{dim}\, \mathcal H \approx 2^{176}$}, we studied the closing of the dissipative gap.
Our findings indicate that the scaling of the gap transforms from a $\Delta \sim 1/L$ \updated{at intermediate Hubbard interaction strength} to a $\Delta\sim 1/L^{3/2}$ behavior once the correlation length exceeds the lattice size, and the local~\gls{BEC} forms, yielding strong evidence for \gls{KPZ}-type scaling of the relaxation dynamics \updated{at mesoscopic system sizes}.
As a consequence, the experimental preparation of a local~\gls{BEC} as suggested in~\cite{Diehl2008} would suffer a superlinear increase of the relaxation time $\tau=1/\Delta$ with the number of lattice sites, when using typical, random initial configurations for the bosons.
However, using~\gls{CLIK-MPS} we were able to show the existence of anomalous relaxation when incorporating additional knowledge of the low\hyp lying Lindbladian eigenmodes.
Specifically, we demonstrated the realization of a~\gls{QME} enabling an exponentially faster relaxation towards the steady state, highlighting the practical relevance of this model as an ideal candidate for the fast, dissipative preparation of a lattice~\gls{BEC} using ultracold atoms.
The possible applications of~\gls{CLIK-MPS} are widespread and range from the numerical investigation of exceptional points \cite{Hatano2019, Minganti2019}, \glspl{DPT} \cite{Minganti2023}, dissipative time crystals \cite{Booker2020, Carollo2024}, the quantum Mpemba effect \cite{Carollo2021,Moroder2024} and metastability \cite{Macieszczak2016} in large\hyp scale, strongly\hyp interacting open quantum systems.
Moreover, combining several complex contours, also spectral regions within the bulk spectrum of the Lindbladian can be targeted, unlocking the investigation of transport and thermalization in driven quantum many\hyp body systems.
Finally, while in this work we considered only a simple analogue of a Markovian embedding, our approach can be immediately extended to the general, non\hyp Markovian case, considering for instance the \gls{HEOM} generator \cite{Debecker2024Spectral, Debecker2024}.
\section*{Acknowledgements}
We are grateful to François Damanet, Baptiste Debecker, John Martin, John Goold and Laetitia Bettmann for enlightening discussions.

PW, US and SP acknowledge support by the Deutsche Forschungsgemeinschaft (DFG, German Research Foundation) under Germany’s Excellence Strategy-426 EXC-2111-390814868.
This work was supported by Grant No. INST 86/1885-1 FUGG of the German Research Foundation (DFG).
MM acknowledges funding from the Royal Society and Research Ireland.
All calculations were performed using the \textsc{SyTen} toolkit \cite{hubig:_syten_toolk,hubig17:_symmet_protec_tensor_networ}.
%
\FloatBarrier
\bibliography{Literature}
%

%
\appendix 
\section{Comparison to Quantum Jumps}
\updated{
To assess the performance of \gls{CLIK-MPS}, we compare it to the \gls{QJ} method \cite{Daley2014}. 
In \gls{QJ}, the evolution of $\hat{\rho}(t)$ is unravelled in multiple stochastic time evolutions of pure states $\ket{\psi_k(t)}$, known as quantum trajectories. The deterministic part of the dynamics is generated by the effective Hamiltonian $\hat H_\mathrm{eff} = \hat{H}-i/2\sum_l \hat L^\dagger_l \hat L^\nodagger_l$, while the stochastic part is given by the applications of jump operators $\hat L_l$ at random times, according to a probability distribution dictated by the norm decrease of $\ket{\psi_k(t)}$ induced by the non-Hermitian $\hat H_\mathrm{eff}$.
\begin{figure}[t!]
    \centering
    \subfloat[\label{submat:subfig:qj:ed}]{
    \includegraphics[width=0.9\columnwidth]{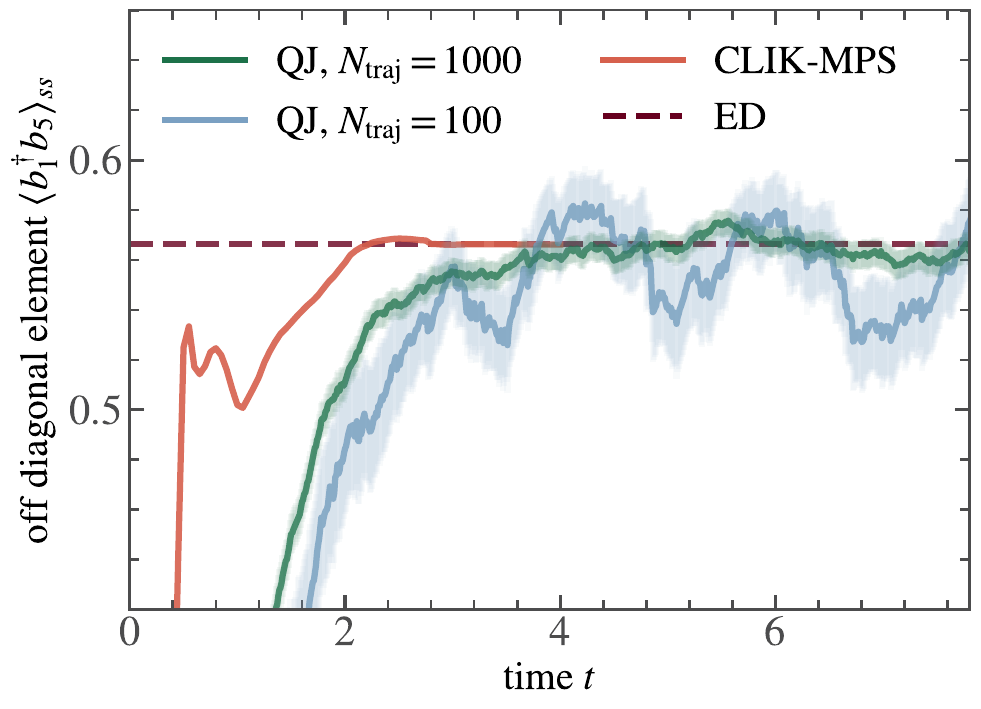}
    }
    \vspace{-
0.6cm}
    \subfloat[\label{submat:subfig:qj:large:system:1}]{
    \includegraphics[width=0.96\columnwidth]{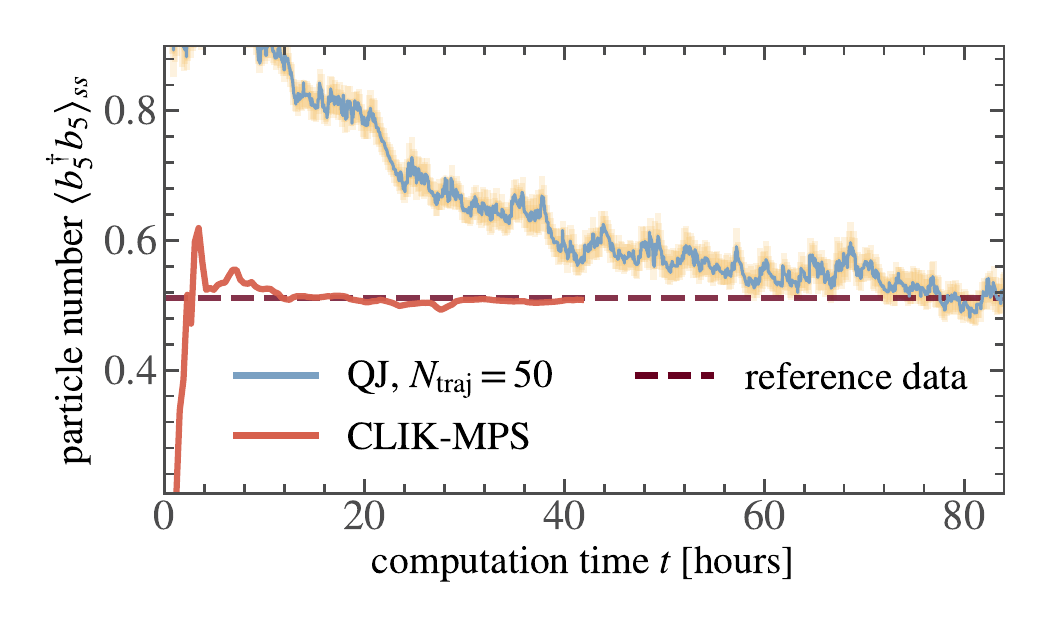}
    }
    \vspace{-
0.8cm}
    \subfloat[\label{submat:subfig:qj:large:system:2}]{
    \includegraphics[width=0.96\columnwidth]{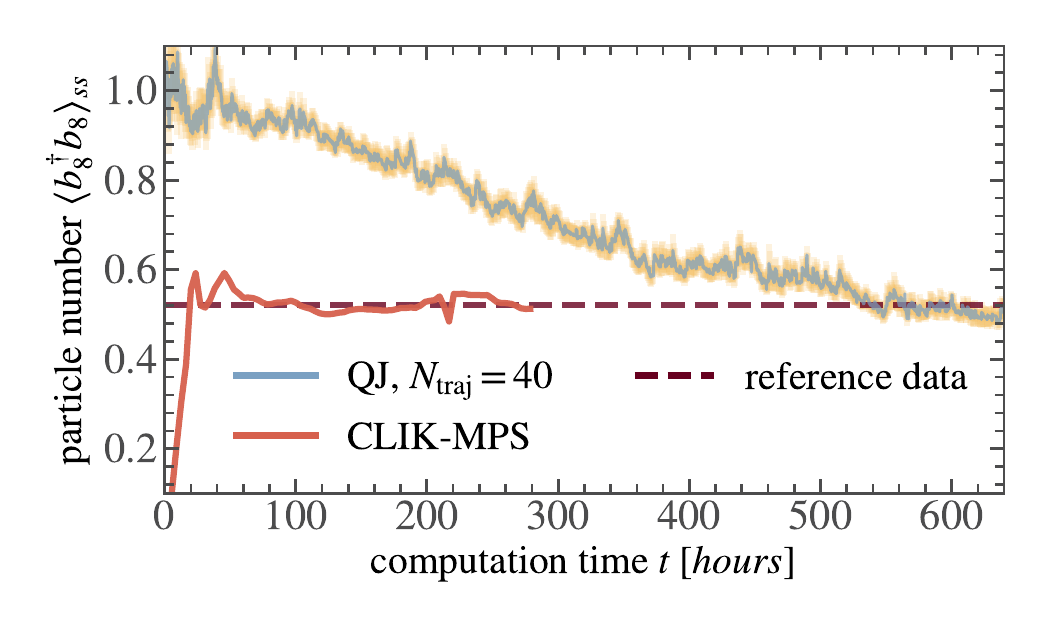}
    }
    \caption{ 
    \updated{Comparison of \gls{CLIK-MPS} to \gls{QJ}. In \cref{submat:subfig:qj:ed} we compare \gls{CLIK-MPS} to \gls{QJ} and \gls{ED} data for a small system of 5 sites and 4 bosons. 
    We used parameters $J=1$, $U=1$ and $\kappa=1$, a timestep of $\mathrm dt =\delta t =0.05$ for \gls{CLIK-MPS} and $\mathrm d t =0.005$ for \gls{QJ}.
    In \cref{submat:subfig:qj:large:system:1} and \ref{submat:subfig:qj:large:system:2} we compare large system data ($L=20$, $N=10$ and $L=32$, $N=16$, respectively) as a function of the runtime. The statistical \gls{QJ} error is shown in shaded yellow around the average value.
    We used parameters $J=1$, $U=0$ and $\kappa=2$, a timestep of $\mathrm d t =0.01$ for \gls{CLIK-MPS} ($\delta t = 0.2$) and $\mathrm d t =0.005$ for \gls{QJ}. For all calculations we used local dimension $d=N+1$. For $L=20$, we fix a maximal bond dimension $\chi=400$, while for $L=32$ we increase to $\chi=600$ for the \gls{CLIK-MPS} calculations. Note that for the calculation of more complicated expectation values a larger bond dimension may be needed.
    }
    }
    \label{app:fig:qj}
\end{figure}
 By averaging over $N_\mathrm{traj}$ trajectories, one can reconstruct expectation value of an observable $\hat{O}$ as
\begin{equation}
\Tr(\hat{O}\hat\rho(t)) \approx \frac{1}{N_\mathrm{traj}}\sum_{k=1}^{N_\mathrm{traj}}\bra{\psi_k(t)}\hat{O}\ket{\psi_k(t)}\,,
\end{equation}}
\updated{where the statistical error is given by the standard deviation of the expectation value across trajectories, divided by $\sqrt{N_\mathrm{traj}}$~\cite{Daley2014}.}

\updated{In~\cref{app:fig:qj}, we compare \gls{CLIK-MPS} to an efficient version of \gls{QJ}, based on an \gls{LSE-TDVP} time evolution \cite{Moroder_diss, Moroder2022}.
%
%
First, in panel a) we simulate a small system with 5 sites and 4 particles, where \gls{ED} data is available. We are interested in the off\hyp diagonal element of the \gls{OBDM}, given by $\langle \hat b_1^\dagger \hat b_5^\nodagger \rangle_\mathrm{ss}$. 
It can be seen that \gls{QJ} needs about $N_\mathrm{traj}=1000$ and reaches convergence at time $t\approx 4$, while \gls{CLIK-MPS} is already converged at $t\approx 2$.
Then in \cref{submat:subfig:qj:large:system:1} and \ref{submat:subfig:qj:large:system:2} we plot the occupation of site $j=L/4$ for the larger system sizes of $L=20$ and $L=32$ at half filling, respectively. To guarantee a fair comparison between \gls{CLIK-MPS} and \gls{QJ}, all calculations were carried out on a dual-socket Intel x86\_64 cluster node with 16 cores per socket (32 physical cores; 64 hardware threads via Hyper-Threading) with 380 GB of memory and x86-64-v4 support. Each run used a single Slurm task with one CPU allocated.
Since \gls{QJ} is trivially parallelizable, we ran each trajectory separately.
Importantly, the shown computational time for \gls{QJ} is that of a \textit{single} trajectory and not the total runtime.
Interestingly, \gls{CLIK-MPS} converges to the reference value (obtained by an accurate long time evolution with low truncation error) substantially faster than \gls{QJ} in both cases. 
Finally, we note that the fact that the number of trajectories needed to converge \gls{QJ} decreases upon increasing $L$ is specific to the parameters we considered: For $U=0$, the steady state is a pure state, up to finite size corrections and each \gls{QJ} trajectory converges to the same steady state, meaning that one needs to compute a single trajectory in the limit $L\to \infty$. For typical mixed steady states, the number of trajectories increases upon increasing the system size when long-range correlations cause the fluctuations of local observables across trajectories to increase with system size~\cite{Daley2014}.
}
\section{Constructing the Krylov subspace}
\label{app:sec:krylov:space}
The starting point of the \gls{CLIK-MPS} framework is to represent the vectorized Lindbladian as a \gls{MPO} \cite{Casagrande2021} and an initial vectorized (potentially mixed) state as a \gls{MPS} \cite{Schollwoeck2011}.
Then a time evolution is performed to obtain a set of vectorized states $\{ \vecket{\rho(t_i)} \}_i$. 
All the details on the efficient implementation of time evolutions can be found in \cref{app:sec:numerical:details}. 
For this set of states we calculate the Gram matrix $\mathbf{M}_{ij} = \vecbraket{\rho(t_i)}{\rho(t_j)}$ and diagonalize it as $\mathbf{M}=\mathbf{U}\mathbf{S}\mathbf{U}^\dagger$. 
Notice that the diagonalization is potentially ill-conditioned due to the high condition number of $\mathbf M$, which is why we use a version of the Block \gls{SVD} for the diagonalization \cite{Stoudenmire2013}.
We discard eigenvalues smaller than a threshold $\epsilon$, which we set to $\epsilon=10^{-14}$ for all calculations in this paper, and advise choosing $\epsilon$ higher than machine precision.
A transformation into an \gls{ONB} is now given by \updated{$\mathbf{X} = \mathbf{U}\mathbf{S}^{-\nicefrac{1}{2}}$} \cite{Paeckel2024}.
The effective Lindbladian in the Krylov subspace is then calculated directly from its matrix elements $\mathbf L_{ij} = \vecbra{\rho(t_i)}\mathcal{\hat L} \vecket{\rho(t_j)}$ on the time-evolved states and transformed into the \gls{ONB} as $\mathbf L^\mathrm{eff} = \mathbf X^\dagger \mathbf L \mathbf X$.
This avoids the numerically unstable procedure of explicitly building an orthonormal basis from $\{ \vecket{\rho(t_i)} \}_i$.
Note that the matrix $\mathbf L^\mathrm{eff}$ does not contain any auxiliary zero eigenvalues that may be introduced by the lower rank of $\mathbf S$ due to the specific definition of  $\mathbf X$.
Diagonalizing $\mathbf L^\mathrm{eff}$ yields the approximate eigenvalues $\lambda_j^{\scriptscriptstyle K}$ and right and left eigenvectors $\mathbf R_j$ and $\mathbf L_j$, respectively.
The Krylov subspace diagonalization procedure is summarized in \cref{app:alg:krylov:subspace}.
These eigenvectors are now given in the Krylov \gls{ONB}.
To write them as direct sums of the initial time-evolved states, we just need to act with $\mathbf X$,
\begin{equation}
    \vecket{r_j^{\scriptscriptstyle K}} = \sum_m \big[ \mathbf X \mathbf R \big]_{mj} \vecket{\rho(t_m)}\; .
    \label{eq:app:krylov:right:eigenvectors}
\end{equation}
Interestingly, this equation reveals a powerful property of the method. 
A central parameter in \gls{MPS} is the bond dimension $\chi$  \cite{Schollwoeck2011}, which determines the maximum amount of correlation the \gls{MPS} can capture. While larger $\chi$ allows for more accurate representations of entangled states, the computational cost of \gls{MPS} operations grows polynomially with $\chi$, making calculations increasingly demanding.
Importantly, the sum of two \gls{MPS} results in a new \gls{MPS} whose bond dimension in general equals the sum of the bond dimensions of the original states.
Thus, if the time-evolved states $\vecket{\rho(t_i)}$ have corresponding bond dimension $\chi\big(\rho(t_i)\big)$, then the maximal bond dimension $\chi_\mathrm{eff}$ reachable within the Krylov space is the bond dimension of a linear combination
\begin{equation}
    \chi_\mathrm{eff}= \sum_{i=1}^{N_{\scriptscriptstyle{K}}} \chi\big(\rho(t_i)\big) \leq N_{\scriptscriptstyle{K}}\max_{i \leq N_{\scriptscriptstyle{K}}} \chi\big(\rho(t_i)\big) \; ,
    \label{eq:app:effective:bdim}
\end{equation}
where $N_{\scriptscriptstyle{K}}$ denotes the number of time-evolved states in the Krylov subspace. 
We term $\chi_\mathrm{eff}$ the \textit{effective Krylov bond dimension}, as it describes the expressibility of the Krylov subspace.
In many cases, $\chi_\mathrm{eff}$ is two orders of magnitude bigger than the ones of the time evolved states due to the linear dependence on $N_{\scriptscriptstyle K}$, allowing for far stronger entangled states.
Crucially, however, such states with large bond dimension never need to be constructed for calculating expectation values or other quantities (see \cref{app:sec:expectations:and:overlaps}) in the Krylov space, rendering the computations very efficient.
This suggests that highly accurate Lindbladian eigenmodes can be constructed from time-evolved states with moderate or even small bond dimensions, provided that the accumulated errors during the dynamics remain controlled.
\begin{algorithm}[H]
\caption{Krylov subspace diagonalization}
\label{app:alg:krylov:subspace}
\begin{algorithmic}[1]
\Procedure{Krylov\_Lindbladian}{$\{\vecket{\rho(t_i)}\}_i$, $\epsilon$}
    \State $\mathbf L \gets$ Lindbladian matrix elements $\vecbra{\rho(t_i)} \hat{\mathcal L} \vecket{\rho(t_j)}$
    \State $\mathbf M \gets$ Gram matrix $\vecbraket{\rho(t_i)}{\rho(t_j)}$
    \State $\mathbf S, \; \mathbf U \gets$ diagonalized Gram matrix $\mathbf M$
    \State $\mathbf S, \; \mathbf U \gets$ discard eigenvalues smaller than $\epsilon$
    \State $\mathbf X \gets \;\updated{\mathbf U\mathbf S^{-1/2}}$
    \Comment{trafo into \gls{ONB}}
    \State $\mathbf L^\mathrm{eff} \gets \; \mathbf X^\dagger \mathbf L \mathbf X$
    \Comment{Action on \gls{ONB}}
    \State $\{\lambda_i, \, \mathbf l^i, \, \mathbf r^i\}_i \gets$ spectral decomp. of $\mathbf L^\mathrm{eff}$
    \State \Return $\{\lambda_i, \mathbf r^i\}_i, \, \mathbf X$
\EndProcedure
\end{algorithmic}
\end{algorithm}

As described in the main text, we not only build the Krylov space out of time-evolved states from one evolution, but also perform two separate time evolutions with different initial states.
Besides a real-time evolution, we perform one along a complex contour, parametrized by $z=t\, \mathrm e^{-\mathrm i\alpha}$ with an angle $\alpha>0$.
This choice of a linearly inclined complex contour changes the weights of the different right eigenmodes during time evolution differently, which is best seen when going into the eigen\hyp decomposition of the Lindbladian, where the time\hyp evolution then reads
\begin{equation}
    \vecket{\rho(z)}=\mathrm e^{\hat{\mathcal L}z}\vecket{\rho_0}= \vecket{\rho_\mathrm{ss}}+\sum_{k=2}^{D^2}\mathrm e^{\lambda_k z}\vecbraket{l_k}{\rho_0}\vecket{r_k} \;.
    \label{app:eq:lindblad:decomposition}
\end{equation}
Consequently, the mode $\vecket{r_k}$ is modulated by an exponential with exponent $\lambda_kz$.
The corresponding exponential dampening is dictated by its real part, which is given by
\begin{equation}
    \xi_k = - \frac{\mathrm{Re} (\lambda_k z)}{t} = 
    -\mathrm{Re}(\lambda_k) \cos(\alpha) - \mathrm{Im}(\lambda_k) \sin(\alpha) \,.
    \label{eq:app:rotate:projection:axis}
\end{equation}
To get more insights into which spectral regions decay more slowly than for a real-time evolution, we can characterize each $\lambda_k$ by $\lambda_k = \vert \lambda_k \vert \mathrm e^{\mathrm i\phi_k}$, yielding
\begin{equation}
\label{suppmat:eq:damping}
    \xi_k = -\mathrm{Re}(\lambda_k) \, \frac{\cos(\alpha-\phi_k)}{\cos(\phi_k)} \, .
\end{equation}
Fixing a value of $\alpha$, eigenvalues whose decay is the same as for $\alpha=0$ satisfy $\cos(\alpha-\phi_k)=\cos(\phi_k)$, which shows that eigenvalues having $\phi_k<\alpha/2+\pi$ are enhanced (i.e. feature lower damping rate $\xi_k< -\mathrm{Re}(\lambda_k)$), while $\phi_k>\alpha/2+\pi$ are suppressed. Consequently, for positive $\alpha$, eigenvalues with $\mathrm{Im}(\lambda_k)\geq0$ are enhanced.
This means that the time-evolved states with positive angle $\alpha$ feature higher weight on the upper half plane than real-time evolved states ($\alpha=0$), where the damping of a mode is dictated by $\mathrm{Re}(\lambda_k)$, which is also the projection of the eigenvalue onto the real axis.
\cref{eq:app:rotate:projection:axis} implies that this picture is slightly changed for the complex contours, and the projection axis has to be rotated by the angle $\alpha$, as shown schematically in \cref{fig:complex:contour}.

\updated{
Note \cref{suppmat:eq:damping} also reveals that the choice of $\alpha$ is not totally free. For instance, if the angle is chosen such that $\alpha> \phi_k - \frac{\pi}{2}$ for some eigenvalue $\lambda_k = \vert \lambda_k \vert \mathrm e^{\mathrm i \phi_k}$, the damping rate $\xi_k$ of the corresponding eigenvalue is negative and the contribution is exponentially enhanced during time evolution. 
In this case, all eigenvalues $\lambda_k$ with $\phi_k<\frac{\pi}{2} + \alpha$ get enhanced, which typically also includes bulk states, destabilizing the time evolution. Additionally, the Krylov space may contain large contributions from bulk states, decreasing the capabilities to resolve the low\hyp lying eigenvalues accurately. 
It is therefore imperative to guarantee that $\alpha \ll \alpha_\mathrm{max} =\min_k \phi_k - \frac{\pi}{2}$.
}

\updated{
We can do so by observing that the time evolution of a traceless initial state $\vecket{\rho}$ shows the behavior
\begin{equation}
\label{app:eq:norm:divergence}
    \vecbraket{\rho(t)}{\rho(t)} \to 
    \begin{cases}
        0 \hspace{0.805cm} \mathrm{if} \; \alpha\leq\alpha_\mathrm{max} \, ,\\
        \infty \hspace{0.62cm} \mathrm{if} \; \alpha>\alpha_\mathrm{max} \, .
    \end{cases}
\end{equation}
where the divergence is exponential. Consequently, by initializing a few time evolutions with different angles $\alpha$ and tracking the norm during a short time evolution, a good approximation to the maximal angle $\alpha_\mathrm{max}$ can be found.
Afterwards, a much lower angle $\alpha\ll \alpha_\mathrm{max}$ is chosen for the time evolution.
}

\updated{
A corresponding algorithm summarizing these ideas is given in \cref{app:alg:find:alpha}. After initializing a (random) traceless state, a short time evolution is carried out for different values of the complex angle $\alpha$. Depending on the evolution of the norm of the state, the angle $\alpha_\mathrm{max}$ can be estimated by checking \cref{app:eq:norm:divergence}.
}
\begin{algorithm}[H]
\caption{\updated{Approximate $\alpha_\mathrm{max}$}}
\label{app:alg:find:alpha}
\begin{algorithmic}[1]
\Procedure{\updated{Get$\alpha_\mathrm{max}$}}{t}
    \State \updated{$\vecket{\rho}$ $\leftarrow$ initialize (random) traceless state}
    \For{\updated{$\alpha$ in grid}}
        \State \updated{$\vecket{\rho(\mathrm e^{-\mathrm i\alpha}t)}\leftarrow$ calculate complex time evolution}
        \If{\updated{Norm is diverging}}
            \State \updated{$\alpha_\mathrm{max}<\alpha$}
        \EndIf
    \EndFor
    \State \Return \updated{$\alpha_\mathrm{max}$}
\EndProcedure
\end{algorithmic}
\end{algorithm}

Notice that the left eigenmode of the steady state to any Lindbladian is given by $\vecket{l_1} = \vecket{\mathds{1}}$. This also includes non-unique steady states, where the identity has to be understood on one specific subspace of the degeneracy, generating strong symmetry. Consequently, initial states (living in a single subspace) that are traceless (and thus unphysical) have vanishing weight on the steady state sector,
\begin{equation}
\label{app:eq:orthogonal:to:ss}
    0 = \mathrm{Tr}\big( \hat \rho_\mathrm{initial}\big) = \vecbraket{\mathds{1}}{\rho_\mathrm{initial}} = \vecbraket{l_1}{ \rho_\mathrm{initial}} \, .
\end{equation}
This property is preserved by the time evolution, and thus (up to truncation errors) the set of time-evolved states will be orthogonal to the steady state. 
If we normalize during time evolution, these states even converge to the slowest decaying mode instead of the steady state 
Note that in the case of the dissipative Bose-Hubbard model, where the particle number conservation enforces a strong $U(1)$ symmetry, we simulate the dynamics in a specific symmetry sector (i.e., at a given particle number).
Note that to generate a Krylov subspace that accurately captures the left Lindbladian eigenmatrices, one needs to perform the time evolution with $\mathcal{L}^\dagger$ (see \cref{eq:lindblad:dagger:decomposition} in the main text).
Then, one can apply \cref{app:alg:krylov:subspace}, with the sole difference being that we include the steady state of $\mathcal{L}^\dagger$, which is the vectorized identity $\vecket{\mathds{1}}$ restricted to a given strong symmetry sector. If there are no conserved quantities, we add the vectorized identity on the whole space. For example, in the case of the Bose Hubbard model considered in the main text, we add the identity restricted to the total particle number sector we fixed through the initial state \cite{westhoff2025fast, Paeckel2019}. This way, we do not introduce contributions from other subsectors into the Krylov space. Since the right eigenmodes of $\hat{\mathcal L}^\dagger$ are actually the left eigenmodes of $\hat{\mathcal L}$, the left eigenmodes are also constructed using \cref{eq:app:krylov:right:eigenvectors}.
\section{Warmup -- optimal initialization}
\label{app:sec:initial:states}
Aside from the maximum evolution time and complex angle, the initial state is the most critical input to the algorithm.
For the method to perform reliably, it must be chosen carefully and according to a systematic strategy.
In this section, we aim to provide physically motivated algorithms to generate good initial states, which are readily generalized to other systems.
The framework presented in this article spans over two different time evolution schemes: the propagation of physical states $\vecket{\sigma^\mathrm{p}_n}$ and traceless states $\vecket{\sigma^\mathrm{u}_n}$, with either $\hat{\mathcal L}$ or $\hat{\mathcal L}^\dagger$.
The efficient initialization of these requires two different optimization protocols, which we discuss below.

First, let us discuss the warmup of $\vecket{\sigma^\mathrm{p}_n}$. At late times $t\gg 1/\mathrm{Re}(\lambda_2)$, the time evolution of an initial state with finite trace will lead to a state close to the steady state.
Since we want to capture the steady state and the slowest decaying modes with this time evolution efficiently, an initial state with a large overlap with modes corresponding to eigenvalues with small modules of their real parts would be favorable. This is implemented by minimizing the real part of the expectation value and variance of the Lindbladian on a subspace spanned by stochastically chosen states.
The resulting routine is summarized in \cref{app:alg:initial:state:1}.
The random subspace is initialized by a product state $\ket{\psi_0} = \ket{n_1, \, n_2, \, \dots\, , n_L}$ on the physical lattice, which is appended to a list $S$. 
In the case of the dissipative Bose-Hubbard model considered here, we found that states having most particles located on the ends of the chain are admissible, since they have a low expectation value.
We emphasize that the effectiveness of a particular initial state strongly depends on the specific structure of the Lindbladian.
Now, $n_\mathrm{samples}-1$ different states are generated stochastically as follows:
In each iteration, a state $\ket{\psi}$ is randomly selected from $S$, along with an index $k$ labeling an occupied site, and a neighboring site $i \in {k \pm 1}$ (note that if $k$ lies at the boundary, only one neighboring site exists).
A new state $\ket{\tilde\psi}$ is built by removing a particle on site $k$ and inserting one at $k+i$ in $\ket{\psi}$. 
The state is added to $S$ if it is not already included, generating a set of $n_\mathrm{samples}$ states that are connected through successive nearest\hyp neighbor hoppings. This is favourable, since the Lindbladian is local and thus finite transition rates between different states will occur. 
Note that for Lindbladians that do not preserve the total particle number, it might be beneficial to also include stochastic single particle loss and injection in addition to hopping.
Once again, we stress that the procedure for sampling basis states is motivated by the specific dissipative Bose-Hubbard Lindbladian that we considered here; for other systems, slight changes might improve the procedure.
A set of vectorized states can then be built by taking the tensor product between states in $S$. 
This yields the set $R=S\otimes S$, containing all $\vecket{\rho} = \ket{\psi} \otimes \ket{\phi}$ with $\ket{\psi}$ and $\ket{\phi}$ in $S$. 
The set $R$ features $n_\mathrm{samples}^2$ different orthonormal states on the vectorized lattice.
The effective Lindbladian action $\mathbf L^\mathrm{eff}$ on $R$ is given by $\mathbf L^\mathrm{eff}_{ij} = \vecbra{\rho_i}\hat{\mathcal{L}}\vecket{\rho_j}$, where $\vecket{\rho_i}, \; \vecket{\rho_j} \in R$.
We diagonalize $\mathbf L^\mathrm{eff}$, pick the lowest lying right eigenvector $\mathbf r^0$ as our initial state and construct its \gls{MPS}\hyp representation by $\vecket{\rho_\mathrm{initial}} = \sum_i \mathbf r^0_i \vecket{\rho_i}$.
\begin{algorithm}[H]
\caption{Compute optimal initial state $\vecket{\sigma^\mathrm{p}_n}$}
\label{app:alg:initial:state:1}
\begin{algorithmic}[1]
\Procedure{Initial\_1}{$\ket{\psi_0}$, $n_\mathrm{samples}$}
    \State $S \gets \big[\ket{\psi_0}\big]$
    \Comment{Initialize with the input state}
    
    \For{$n = 1$ to $n_\mathrm{samples}-1$}
        \State $\ket{\psi} \gets$ randomly drawn from $S$
        \State $k \gets$ random occupied site in $\ket{\psi}$
        \State $i \gets$ random number from $\pm1$
        \State $\ket{\tilde{\psi}} \gets$ $\hat b_{k+i}^\dagger \hat b_k^\nodagger \ket{\psi}$
        \If{$\ket{\tilde \psi} \notin S$}
            \State Append $\ket{\tilde \psi}$ to $S$
        \Else
            \State Go back to line 4
        \EndIf
    \EndFor

    \State $R \gets \big[\ket \psi \otimes \ket\phi \; \mathrm{for}\; \ket \psi, \, \ket\phi \; \mathrm{in} \;S\big]$

    \State $\mathbf L^\mathrm{eff} \gets$ $\vecbra{\rho_i}\hat{\mathcal{L}}\vecket{\rho_j}$ for $\vecket{\rho_i}, \; \vecket{\rho_j}$ in $R$
    \State $\lambda_0, \mathbf r^0 \gets$ lowest eigenvalue and right eigenvector of $\mathbf L^\mathrm{eff}$
    \State $\vecket{\rho_\mathrm{initial}} \gets$ build $\sum_i \mathbf r^0_i \vecket{\rho_i}$ for $\vecket{\rho_i}$ in $R$

    \If{$\big|\vecbraket{\mathds{1}}{\rho_\mathrm{initial}}\big| < \epsilon_\mathrm{trace}$}
        \State Go back to line 3
    \EndIf

    \State \Return $\vecket{\rho_\mathrm{initial}},\; S $
\EndProcedure
\end{algorithmic}
\end{algorithm}
\vspace{0.4cm}
\noindent To ensure it is a traceful state, we calculate its trace norm $\vecbraket{\mathds{1}}{\rho_\mathrm{initial}}$ and discard the state and repeat the procedure if it happens to be smaller than some threshold $\epsilon_\mathrm{trace}$.
 With some minor adjustments, this algorithm can also be used to find optimal initializations for $\vecket{\sigma^\mathrm{u}_n}$, i.e. \textit{traceless} states with low expectation value and variance with the Lindbladian (see also \cref{app:eq:orthogonal:to:ss}). 
This constraint is elegantly incorporated by choosing only traceless basis states for the subspace $R$, making every spanned state in the subspace immediately traceless. 
The corresponding pseudocode is given in \cref{app:alg:initial:state:2}.
As mentioned, the basic idea stays the same, but the subspace in which the optimal state is searched for (spanned by the randomly chosen states) should only contain traceless states. 
Vectorized states of the form $\ket \psi\otimes\ket \phi$ with $\ket\psi, \; \ket\phi \in S$ if $\ket\psi\neq\ket\phi$ are already traceless and can be appended to $R$.
However, $\ket\psi\otimes\ket\psi$ has trace one, so we may instead take $\ket\psi\otimes\ket\psi-\ket\phi\otimes\ket\phi$ into $R$. 
Notice that the procedure leads to $n_\mathrm{samples}^2-1$ states in $R$ living on the vectorized lattice. 
Different from \cref{app:alg:initial:state:1}, these states are not orthonormal, and we need to build an \gls{ONB} from them. 
This can be done by employing \cref{app:alg:krylov:subspace} initialized with the set of states $R$ and some threshold for the retained eigenvalues $\epsilon$. 
Notice that due to the simple structure of the states in $R$, the Gram matrix may also be constructed explicitly without calculating any overlaps.
\cref{app:alg:krylov:subspace} returns the eigenvalues and corresponding right eigenvectors $\{ \lambda_i, \, \mathbf r^i\}_i$ of the effective Lindbladian in the Krylov space, and the transformation into the Krylov\hyp \gls{ONB}.
The right eigenvector $\mathbf r^0$ to the highest eigenvalue is now the state with the lowest real part of the Lindbladian expectation in the subspace, and thus the optimal initial state. 
Back in the \gls{MPS} representation, this state is constructed as $\vecket{\rho_\mathrm{initial}}=\sum_i \big[ \mathbf X \mathbf r^0\big]_i \vecket{\rho_i}$.
\begin{algorithm}[H]
\caption{Compute optimal initial state $\vecket{\sigma^\mathrm{u}_n}$}
\label{app:alg:initial:state:2}
\begin{algorithmic}[1]
\Procedure{Initial\_2}{$\ket{\psi_0}$, $n_\mathrm{samples}$}
    \State $S \,\, \gets$ do lines 2 -- 13 from \cref{app:alg:initial:state:1}

    \State $R \; \gets \big[\ket \psi \otimes \ket\phi \; \mathrm{for}\; \ket \psi, \, \ket\phi \; \mathrm{in} \;S \; \mathrm{if} \; \ket{\psi}\neq\ket{\phi} \big]$
    \State $D \,\gets \big[ \ket{\psi}\otimes \ket{\psi} \; \mathrm{for} \; \ket{\psi} \; \mathrm{in} \; S \big]$
    \For{$n=1$ to $n_\mathrm{sites}-1$}
        \State Append $\vecket{\delta_n}-\vecket{\delta_{n+1}}$ to $R$ for $\vecket{\delta_n}, \; \vecket{\delta_{n+1}}$ in $D$
    \EndFor
    \State $\{\lambda_i, \mathbf r^i\}_i, \, \mathbf X \gets$  do \cref{app:alg:krylov:subspace} with states $R$
    \State $\lambda_0, \mathbf r^0 \gets$ highest eigenvalue and right eigenvector
    \State $\vecket{\rho_\mathrm{initial}} \gets$ build $\sum_i \big[ \mathbf X \mathbf r^0\big]_i \vecket{\rho_i}$ for $\vecket{\rho_i}$ in $R$

    \State \Return $\vecket{\rho_\mathrm{initial}} $
\EndProcedure
\end{algorithmic}
\end{algorithm}
Although \cref{app:alg:initial:state:1} and \ref{app:alg:initial:state:2} return a good initial state in the stochastically chosen subspaces, there is no guarantee that there might not be a better subspace in the first place. 
This problem can be resolved by introducing an iterative extension to the methods, shown in \cref{app:alg:iterative:initial}.
Instead of directly building the whole subspace in one step, we now iteratively find the full subspace by successively adding a few random states and optimizing this smaller subspace in each step. Specifically, given a state list $S$ containing $n$ states, we stochastically generate a new state list $S'$ by adding two states through hopping to $S$ and evaluating the lowest eigenvalue. The same process is repeated a couple of times, and afterwards we chose the one set $S'$ which gave rise to the lowest eigenvalue. This leaves us with a set $S$ of size $n+2$, and the process may be repeated.
This method lowers the expectation value systematically in every iteration, and it can be ended either if a desired Krylov space dimension is attained, or a threshold bond dimension or Lindbladian expectation value is reached.
The former is implemented via the choice of $n_\mathrm{samples}$, which bounds the Krylov space dimension by $n_\mathrm{samples}^2$, while the latter is implemented as an escape criterion.

In \cref{fig:algorithm:warmup}, such an iterative procedure is shown for the dissipative Bose-Hubbard model with 10 sites and 5 particles. 
The warmup was performed until $n_\mathrm{states}=16$ for both a traceless and a physical initial state. 
In every iteration, the expectation value is lowered, which shows the benefit of iteratively applying \cref{app:alg:initial:state:1} and \ref{app:alg:initial:state:2}.
The corresponding bond dimensions (see inset) increase with $n_\mathrm{states}$ to about $\chi =20$ (physical) and $\chi =60$ (traceless). 
Since in every iteration only 2 basis vectors are added, not all matrix elements have to be calculated, since some can be reused from the earlier iteration.
\begin{algorithm}[H]
\caption{Iteratively compute optimal initial states}
\label{app:alg:iterative:initial}
\begin{algorithmic}[1]
\Procedure{Iterate\_Initial}{$\ket{\psi_0}$, $n_\mathrm{samples}$}    
    \For{$n = 1$ to $n_\mathrm{samples}/2$}
        \For{$k = 1$ to $3 n_\mathrm{sites}$}
            \State $\vecket{\rho(k)}, \, S(k) \gets$ Do \cref{app:alg:initial:state:1} or \ref{app:alg:initial:state:2}  with $(S, \;2)$
        \EndFor
        \State $k_\mathrm{max} \gets \; \mathrm{argmax}_k \,\mathrm{Re} \vecbra{\rho(k)}\hat{\mathcal{L}}\vecket{\rho(k)}$
        \State $\vecket{\rho_\mathrm{initial}}, \; S \gets \, \vecket{\rho(k_\mathrm{max})}, \; S(k_\mathrm{max})$
        \If{$\chi(\rho_\mathrm{initial})>\chi_\mathrm{max}$}
            \State Break
        \EndIf
    \EndFor

    \State \Return $\vecket{\rho_\mathrm{initial}} $
\EndProcedure
\end{algorithmic}
\end{algorithm}

For the time evolution with $\mathcal{\hat L}^\dagger$, the search for an optimal initial state is not as easily physically motivated.
Unfortunately, we do not know the structure of the left eigenvector corresponding to the steady state, and thus orthogonalizing against it is not doable.
Minimizing the expectation value of $\hat{\mathcal L}^\dagger$ is nevertheless useful, and we thus just use the same algorithms as for the time evolution with $\hat{\mathcal L}$. 
The sole adjustment is that the effective Lindbladian $\mathbf L^\mathrm{eff}$ has to be replaced by the adjoint effective Lindbladian, $\vecbra{\rho_i}\hat{\mathcal L}^\dagger\vecket{\rho_j}$.
Thus, for time evolutions with zero complex angle $\alpha = 0$, we use \cref{app:alg:initial:state:1}, whereas for finite angles, \cref{app:alg:initial:state:2} is employed (both within the iterative extension described in \cref{app:alg:iterative:initial}).
\section{Computing expectation values and overlaps}
\label{app:sec:expectations:and:overlaps}
\begin{figure}
    \centering
    \includegraphics[width=0.9\columnwidth]{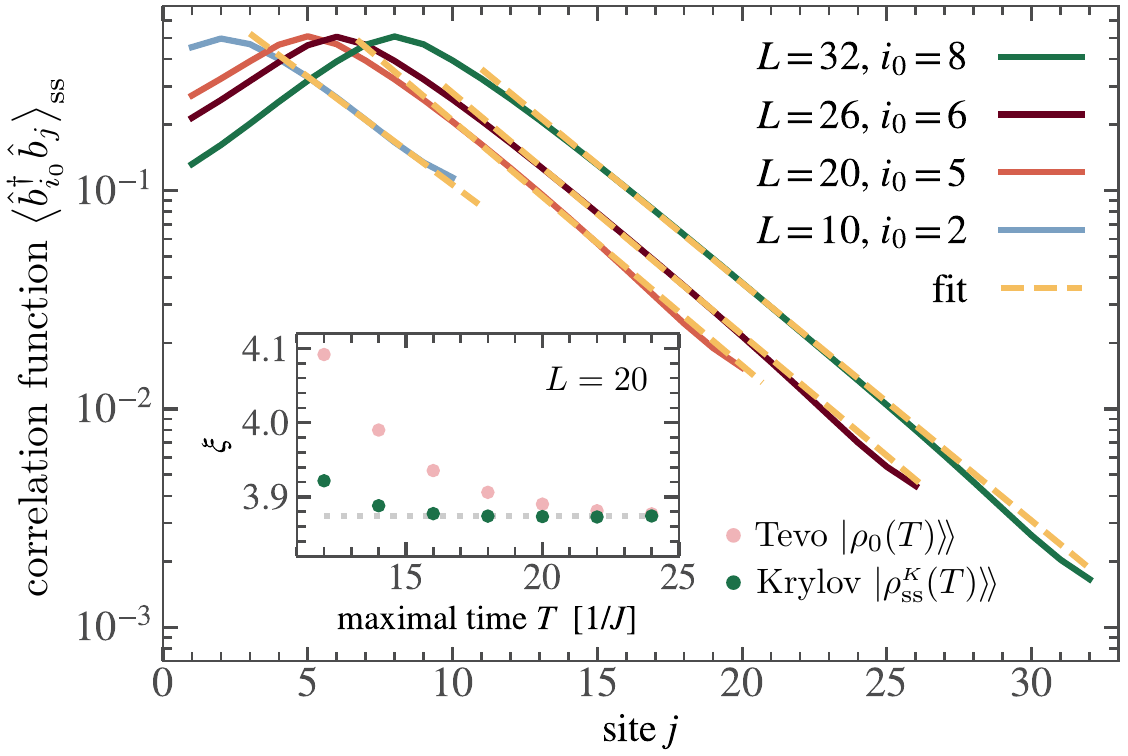}
    \caption{
    Steady state from large system calculations. 
    Correlation function for 10 (blue), 20 (orange), 26 (red) and 32 (green) sites depending on the second site index $j$ at fixed first index $i_0=L/4$ at $U=2J$. We fit the tail of the correlation function exponentially (yellow), and recover the correlation length $\xi$ of the many-body mixed state. The Krylov spaces used time-evolved states until 6, 18, 22 and 32 with a sampling spacing of 0.1, 0.2, 0.4 and 0.4 for the different system sizes respectively.
    Inset: Correlation length $\xi$ depending on the maximal time $T$ for 20 sites. We compare the \gls{CLIK-MPS} steady state $\vecket{\rho_\mathrm{ss}^{\scriptscriptstyle K}(T)}$ with the last time evolved state $\vecket{\rho_0(T)}$.
    For all data shown here, we chose $J=1$, $\kappa=2J$, a local dimension $d=N+1$ and for $L\leq 20$ we set the maximal bond dimension $\chi =2400$, while for $L> 20$ we have $\chi=1400$.
    }
    \label{app:fig:correlation:function:fit}
\end{figure}
In most applications, also throughout this paper, the explicit \gls{MPS} representation of the steady state and slowly decaying modes is not of primary interest; rather, one typically focuses on specific expectation values or overlaps with them.
In \gls{CLIK-MPS} it is actually possible to avoid constructing the \gls{MPS} representation of the approximate eigenvectors; It suffices to compute overlaps and expectation values with the original time-evolved states.
Assume we are interested in the expectation of operator $\hat A$ (for instance, some particle-particle correlator $\hat A_{ij} = \hat b^\dagger_i \hat b^\nodagger_j$ in the main text).
As shown in \cref{eq:app:krylov:right:eigenvectors}, the unnormalized steady state can be represented easily in the original basis.
Thus, if we define the action of $\hat A$ on the Krylov basis as $\mathbf A^\mathrm{eff}_{n} = \vecbra{\mathds{1}}\hat A\vecket{\rho(t_n)}$ and the normalization $\mathbf N_n=\vecbraket{\mathds{1}}{\rho(t_n)}$, it is straightforward to calculate expectation values
\begin{equation}
\label{app:eqn:expectation:values}
    \mathrm{Tr}\big(\hat A \hat \rho_\mathrm{ss}\big) = \frac{\big[\mathbf A^\mathrm{eff} \mathbf{XR}\big]_1}{\big[ \mathbf{NXR} \big]_1}\; .
\end{equation}
 Note that the vectorized identity is in general well-described by a \gls{MPS} with low bond dimension; in all calculations performed in the scope of this article, it did not exceed $\chi\sim 10^3$.

Since the time evolution with a nonzero complex angle is not Hermiticity preserving, the steady state $\vecket{r_1^{\scriptscriptstyle K}}$ might have some non\hyp Hermitian artifacts due to the finite $\alpha$ contributions, which are also present in \cref{app:eqn:expectation:values}. 
This issue can be resolved by instead considering the hermitian part of the density matrix, given by $\vecket{r_{1, \mathrm{herm}}^{\scriptscriptstyle K}}=\nicefrac{1}{2} \,\big(\vecket{r^{\scriptscriptstyle K}_1}+\vecket{{r_1^{\scriptscriptstyle{K}}}^\dagger}\big)$, where $\vecket{{r_1^{\scriptscriptstyle{K}}}^\dagger}$ denotes the vectorization of $\big[\hat r_1^{\scriptscriptstyle K}\big]^\dagger$.
Expectation values of $\hat A$ with this hermitian steady state can be calculated easily within the Krylov space. Besides $\mathbf A^\mathrm{eff}$ we define $\mathbf A^\mathrm{eff, *}_n = \vecbra{\mathds{1}}\hat A^\dagger\vecket{\rho(t_n)}$, and find
\begin{equation}
    \mathrm{Tr}\big(\hat A\hat{\rho}_\mathrm{ss}\big) = \frac{1}{2\big[ \mathbf{NXR} \big]_1}\Big(\big[\mathbf A^\mathrm{eff} \mathbf{XR}\big]_1 + \big[\mathbf A^\mathrm{eff, *} \mathbf{XR}\big]_1^*\Big) \, .
    \label{app:eq:hermitian:observables}
\end{equation}
This equation eliminates non\hyp Hermitian artifacts and we use it throughout the main text when calculating the two-point correlator.
Similarly, overlaps of arbitrary states with the Krylov eigenvectors can be calculated.
If we are interested in the overlap of some vectorized density matrix $\vecket{\rho}$ with the Krylov eigenmode $\vecket{r_j^{\scriptscriptstyle K}}$, it suffices to calculate the overlap vector $\mathbf O_n = \vecbraket{\rho}{\rho(t_n)}$ to obtain
\begin{equation}
    \vecbraket{\rho}{r_j^{\scriptscriptstyle K}} = \big[\mathbf {OXR}  \big]_j \,.
\end{equation}
Crucially, we did not need to construct the explicit \gls{MPS} representations of any eigenvector in the Krylov space, which drastically reduces computational complexity.
Note that it is enough to describe overlaps with right eigenvectors, as left eigenvectors of $\hat{\mathcal L}$ are calculated in \gls{CLIK-MPS} by considering time evolution with $\hat{\mathcal L}^\dagger$, whose right eigenvectors are the left eigenvectors of $\hat{\mathcal L}$.
Our framework gives access to highly accurate eigenvalues and eigenstates on the upper half plane. 
Due to the Lindbladian being hermitian-preserving, the spectrum is symmetric around the real axis, and eigenvalues on the lower half plane can be trivially obtained by complex conjugation.
Eigenvectors on the lower half plane can also be found by hermitian conjugation of the ones on the upper plane.
Overlaps with them are nevertheless easy to calculate within \gls{CLIK-MPS}, we only need the useful identity
\begin{equation}
    \vecbraket{r_i^\dagger}{\rho} = \vecbraket{r_i^\nodagger} {\rho}^* \,,
\end{equation}
which is easily proven in the density matrix picture using the properties of the trace and $\hat{\rho}$ being hermitian. 
Thus, the overlap with  $\vecket{r_i^\dagger}$ is easily computed within \gls{CLIK-MPS}.
An application of the observable calculation in the steady state of the Krylov space can be seen in \cref{app:fig:correlation:function:fit}. As in the main text, we are interested in the correlation function $\big\langle\hat b_i^\dagger\hat b_j^\nodagger\big\rangle_\mathrm{ss}=\mathrm{Tr}\big(\hat b_i^\dagger\hat b_j^\nodagger \hat \rho_\mathrm{ss}\big)$. For the steady state, we use the approximate steady state obtained via \gls{CLIK-MPS}, and label it with the maximal Krylov time $T$, giving $\vecket{\rho_\mathrm{ss}^{\scriptscriptstyle K}(T)}$. We fix the first index at $i_0=L/4$ to avoid influence from the boundaries.
The observables can be easily obtained using \cref{app:eq:hermitian:observables}.
We find that the correlation function decays exponentially, as expected within the mean field theory \cite{Diehl2008}. The corresponding correlation length $\xi$ can be fitted from the tail of the correlation function using an exponential fitting function, where we again dismiss the boundary sites. 
Convergence of the correlation length can be monitored by building the Krylov subspace spanning over states up to a shorter maximal time $T$ and comparing $\xi$. A detailed discussion of the convergence analysis can be found in \cref{app:sec:convergence:large:systems}.
For all correlation lengths shown in the main text in \cref{fig:steady-state:correlation-lengths}, the same calculations were performed.

\section{Exploiting symmetries}
\label{sec:app:discrete:symmetries}
Exploiting symmetries is a key strategy for improving the efficiency of observable calculations.
Assume the Lindbladian at hand possesses a \textit{discrete symmetry}, which comes with a generator $\hat{\mathcal Q}$ and a unitary $\hat{\mathcal U}$ connected via
\begin{equation}
    \hat{\mathcal U} = \mathrm{exp}\big( -\mathrm 2\pi i\,\hat{\mathcal Q} \big) \; .
\end{equation}
In most practical cases, including the example of $\mathds{Z}_{N_S}$\hyp symmetries, the vectorized unitary $\hat{\mathcal U}$ is constructed from another unitary $\hat U$ acting on the physical and auxiliary space $\hat{\mathcal U} = \hat U\otimes \hat U^*$. The generator $\hat Q$ of $\hat U$ is thus given by $\hat{\mathcal Q} = \hat Q \otimes \hat{\mathds 1} - \hat{\mathds 1} \otimes \hat Q$.
In the case of $\mathds{Z}_{N_S}$ symmetries, $\hat Q$ has spectrum $j/N_S$ for $0\leq j < N_S$. Notice that this directly fixes the spectrum of $\hat{\mathcal Q}$.
If the Lindbladian now exhibits this symmetry -- i.e., $[\hat{\mathcal L}, \hat{\mathcal U}] = 0$ -- it decomposes into $N_S$ blocks, each characterized by a fixed transformation behavior and labeled by the eigenvalues $q$ of $\hat{\mathcal{Q}}$.
All Lindbladian eigenvectors $\vecket{l_j}, \, \vecket{r_j}$ are thus also eigenvectors of $\hat{\mathcal U}$ and live in a fixed symmetry sector $q$.
Consequently, we would like to use not the approximate Krylov eigenvector as steady state, but instead its projection into the subspace it belongs to due to the symmetry. We label this symmetry sector by $q_0$.
In mathematical terms, this translates to 
\begin{equation}
    \vecket{\rho_\mathrm{ss}^{\scriptscriptstyle K, q_0}} = \hat{\mathcal P}_{q_0} \vecket{\rho_\mathrm{ss}^{\scriptscriptstyle K}} \, ,
\end{equation}
where $\hat{\mathcal P}_{q_0}$ is the projector onto the $q_0$\hyp sector.
Crucially, this projector can be written as a polynomial in $\hat{\mathcal U}$,
\begin{equation}
\label{app:eq:projector:for:u}
    \hat{\mathcal P}_{q_0} = \frac{1}{N_S} \sum_{n=0}^{N_S-1} \mathrm{e}^{2\pi\mathrm i q_0 n} \, \hat{\mathcal U}^n \,.
\end{equation}
The key question that remains is identifying the $q_0$ sector in which the steady state resides.
Interestingly, one can show that the steady state is always in the $q_0=0$ sector.
Since the steady state $\vecket{\rho_\mathrm{ss}}$ has a finite trace, it contains a component lying in the 0\hyp sector due to
\begin{align}
    \vecbra{\mathds 1} \hat{\mathcal P}_{q_0} &\vecket{\rho_\mathrm{ss}} = \frac{1}{N_S}\sum_{n=0}^{N_S-1} \mathrm{e}^{2\pi\mathrm i q_0 n} \vecbra{\mathds 1} \hat{\mathcal U}^n\vecket{\rho_\mathrm{ss}} \nonumber\\
    & = \frac{\vecbraket{\mathds 1}{\rho_\mathrm{ss}}}{N_S}\sum_{n=0}^{N_S-1} \mathrm{e}^{2\pi\mathrm i q_0 n} = \vecbraket{\mathds 1}{\rho_\mathrm{ss}} \delta_{q_0=0} \,.
\end{align}
Due to the fact that $\hat{\mathcal U} \vecket{\mathds 1}= \vecket{\mathds 1}$, which is seen by going back into the matrix picture, $\hat{\mathcal U} \vecket{\mathds 1} \to \hat U^\dagger \mathds 1\hat U^\nodagger = \mathds 1$. The only possible sector eligible for the steady state is thus the $q_0=0$ sector.
Now, we can safely insert the symmetrized approximate steady state $\hat{\mathcal P}_{q_0=0}\vecket{\rho_\mathrm{ss}^{\scriptscriptstyle K}}$ into the observable
\begin{align}
    \mathrm{Tr}\big( \hat A \hat \rho&_\mathrm{ss} \big) = \vecbra{\mathds 1} \hat A \hat{\mathcal P}_{q_0=0}\vecket{\rho_\mathrm{ss}^{\scriptscriptstyle K}} \nonumber \\
    &= \frac{1}{N_S}\sum_{n=0}^{N_S-1} \vecbra{\mathds 1} \hat{\mathcal U}^n(\hat{\mathcal U}^\dagger)^n \hat A \hat{\mathcal U}^n\vecket{\rho_\mathrm{ss}^{\scriptscriptstyle K}} \,.
\end{align}
The expression is brought into its final form by defining the overlaps $\mathbf A^\mathrm{eff}_{n,k} = \vecbra{\mathds 1}(\hat{U}^\dagger)^n \hat A \, \hat{U}^n\vecket{\psi_k}$, which yields
\begin{equation}
\label{app:eq:symmetry:conserving:observable}
    \mathrm{Tr}\big( \hat A \hat \rho_\mathrm{ss} \big) = \frac{1}{N_S\big[ \mathbf{NXR} \big]_1} \sum_{n=0}^{N_S-1} [\mathbf A^\mathrm{eff} \mathbf{XR}\big]_{n1} \,.
\end{equation}
Computationally, this is equivalent to calculating \cref{app:eqn:expectation:values} $N_S$\hyp  times. The crucial advantage we gain by avoiding applying the projector directly onto $\vecket{\rho_\mathrm{ss}^{\scriptscriptstyle K}}$, is that in most use cases the operator $\hat A$ is much simpler than the states in the Krylov space, and $(\hat{\mathcal U}^\dagger)^n \hat A \hat{\mathcal U}^n$ can be calculated exactly.
Notice that \cref{app:eq:symmetry:conserving:observable} can be combined with the cancellation of non\hyp Hermitian artifacts similarly as in \cref{app:eq:hermitian:observables}, if we also define $\mathbf A^\mathrm{eff, *}_{n,k} = \vecbra{\mathds 1}(\hat{U}^\dagger)^n \hat A^\dagger \, \hat{U}^n\vecket{\psi_k}$ we end up with the final expression
\begin{equation}
\label{app:eq:symmetry:conserving:observable:and:herm}
    \mathrm{Tr}\big( \hat A \hat \rho_\mathrm{ss} \big) = \frac{\sum_{n=0}^{N_S-1}\Big( [\mathbf A^\mathrm{eff} \mathbf{XR}\big]_{n1} + [\mathbf A^\mathrm{eff, *} \mathbf{XR}\big]_{n1}^*\Big)}{2N_S\big[ \mathbf{NXR} \big]_1} \,.
\end{equation}
In the specific case of the dissipative Bose\hyp Hubbard model considered in the main text, there is an additional $\mathds Z_2$ symmetry, given by an inversion of the lattice around the middle site(s). The corresponding unitary $\hat U_\mathrm{inv}$ has the action $\hat U_\mathrm{inv}^\dagger \hat b_j \hat U_\mathrm{inv}^\nodagger = \hat b_{L+1-j}$ and its vectorized version is given by $\hat{\mathcal U}_\mathrm{inv} = \hat U_\mathrm{inv}\otimes\hat U_\mathrm{inv}^*$.
We are mainly interested in the two point correlators $\langle\hat b_j^\dagger\hat b_k^\nodagger\rangle_\mathrm{ss}$, which means we chose $\hat A_{j,k} = \hat b_j^\dagger\hat b_k^\nodagger$ and consequently $\hat{\mathcal U}_\mathrm{inv}^\dagger \hat A_{j,k} \,\hat{\mathcal U}_\mathrm{inv} = \hat b_{L+1-j}^\dagger\hat b_{L+1-k}^\nodagger=\hat A_{L+1-j,L+1-k}$. 
Thus, the resulting computations based on \cref{app:eq:symmetry:conserving:observable} have the same numerical complexity as \cref{app:eqn:expectation:values}.
\updated{
The main bottleneck in generalizing the expressions \cref{app:eq:symmetry:conserving:observable} and (\ref{app:eq:symmetry:conserving:observable:and:herm}) to continuous symmetries, is the explicit expression of the projector $\hat{\mathcal P}$ in terms of the unitary $\hat{\mathcal U}$. Indeed, in the case of a $U(1)$\hyp symmetry, the generator has countably infinite eigenvalues, the unitary depends continuously on a parameter $a$ and \cref{app:eq:projector:for:u} takes on the very familiar form of a Fourier series
\begin{equation}
    \hat{\mathcal P}_{n} = \frac{1}{2\pi} \int_0^1 \mathrm da \;  \mathrm{e}^{2\pi\mathrm i a n} \, \hat{\mathcal U}(a) \,.
\end{equation}
Now, performing the same steps as before yields
\begin{equation}
    \mathrm{Tr}\big( \hat A \hat \rho_\mathrm{ss} \big) = \vecbra{\mathds 1} \hat{\mathfrak A} \vecket{\mathds \rho_\mathrm{ss}^{\scriptscriptstyle K}}\; , \hspace{0.3cm} \hat{\mathfrak A} = \frac{1}{2\pi} \int_0^1 \mathrm da \; \hat{U}^\dagger (a) \hat A \hat{U}(a) \; .
    \label{app:eq:symmetry:conserving:observable:continuous}
\end{equation}
This result is similar to \cref{app:eq:symmetry:conserving:observable} and in spirit, only the sum is changed for an integral.
This approach is readily generalized to other continuous symmetries in the same fashion.
}

\updated{
Also the derivation of \cref{app:eq:symmetry:conserving:observable}, (\ref{app:eq:symmetry:conserving:observable:and:herm}) and (\ref{app:eq:symmetry:conserving:observable:continuous}) solely rely on the fact that $\hat{\mathcal U} \vecket{\mathds 1} = \vecket{\mathds 1}$ and in particular the explicit form of the generator of time evolution is irrelevant, which is particularly useful for the generalization of the framework to non\hyp Markovian settings.
}

\section{Comparing to ED\hyp eigenvectors}
\label{app:sec:ed:comparison}
After the diagonalization in the Krylov subspace, we get access to a matrix $\mathbf R$, encoding all right eigenvectors, and \cref{eq:app:krylov:right:eigenvectors} tells how to write the eigenvectors in the basis of the original time evolved states.
To assess the quality of the approximation, we can compare to \gls{ED} diagonalization of the Lindbladian written as a matrix in some \gls{ONB} $\{\vecket{e_j}\}_j$.
This yields the exact right eigenvectors $\mathbf R_j^{\scriptscriptstyle E}$ written in the \gls{ONB}.
to bring the two different bases together, we represent the time-evolved states used in \gls{CLIK-MPS} in the \gls{ONB} used for \gls{ED} by $\mathbf E_{mj}=\vecbraket{e_m}{\rho(t_j)}$, making it possible to compute the overlap
\begin{equation}
    \vecbraket{r^\nodagger_j}{r_j^{\scriptscriptstyle K}} = \big[{\mathbf R^{\scriptscriptstyle E}}^\dagger\mathbf{ EXR}\big]_{jj} \; ,
    \label{app:ed:overlap}
\end{equation}
where $\vecket{r_j}$ is the $j$\hyp th exact right eigenvector of the Lindbladian. 
\cref{app:ed:overlap} is best understood from right to left: $\mathbf X$ transforms from the Krylov\hyp \gls{ONB} to the original time evolved states, while $\mathbf E$ maps them onto the basis used for \gls{ED}. 
Then the overlap is calculated by acting with the right eigenvectors in this basis from the left.
We use \cref{app:ed:overlap} to calculate overlaps with the exact right and left eigenvectors in \cref{subfig:mps:ed:comparison:reigenvectors} and \ref{subfig:mps:ed:comparison:leigenvectors} in the main text.
\section{Quantifying accuracy beyond ED comparisons}
\label{app:sec:convergence:large:systems}
When considering small systems, the accuracy of the \gls{CLIK-MPS} framework can be tested straight away as outlined in \cref{app:sec:ed:comparison}. 
Instead, for large-scale systems lacking exact reference data, it becomes imperative to develop alternative checks to assess the precision of the computed quantities.

Our main objective is to quantify the deviations from the exact steady state when using \gls{CLIK-MPS}.
We formulate the following theorem: \newline

Denote by $\vecket{\rho_\mathrm{ss}^{\scriptscriptstyle{K}}}$ the approximate steady state from \gls{CLIK-MPS}. Then we have
\begin{equation}
\big|\vecbraket{\rho_\mathrm{ss}^{\scriptscriptstyle{K}}}{\rho_\mathrm{ss}^\nodagger}\big|^2 \geq 1 - \frac{\vecbra{\rho_\mathrm{ss}^{\scriptscriptstyle{K}}} \mathcal{\hat L}^\dagger\mathcal{\hat L}^\nodagger \vecket{\rho_\mathrm{ss}^{\scriptscriptstyle{K}}}}{|\sigma_2|^2} \; .
\end{equation}
For the proof, notice that $\mathcal{\hat L}^\dagger \mathcal{\hat L}$ is hermitian with non-negative spectrum. 
Furthermore, the steady state is its only eigenstate with eigenvalue 0.
The gap of this hermitian operator is given by $\mu_2 = |\sigma_2|^2$, where $\sigma_2$ is the second lowest singular value of $\mathcal{\hat L}$.
We may diagonalize $\mathcal{\hat L}^\dagger \mathcal{\hat L}$, denote its increasingly ordered eigenvalues by $\mu_k$ and its eigenvectors by $\vecket{v_k}$.
Going into the eigenbasis we find $\vecket{\rho_\mathrm{ss}^{\scriptscriptstyle{K}}} = \sum_m a_m \vecket{v_m}$ and
\begin{equation}
    \vecbra{\rho_\mathrm{ss}^{\scriptscriptstyle{K}}} \mathcal{\hat L}^\dagger\mathcal{\hat L}^\nodagger \vecket{\rho_\mathrm{ss}^{\scriptscriptstyle{K}}} = \sum_{m\geq 2} \mu_m |a_m|^2 \; .
\end{equation}
Combining this with the normalization constraint $1=\sum_m |a_m|^2$ we find the bound
\begin{equation}
    |a_1|^2 = 1 - \sum_{m\geq2}|a_m|^2 \geq 1 - \frac{1}{\mu_2}\sum_{m\geq2}\mu_m|a_m|^2 \; ,
\end{equation}
where we used that $\mu_m/\mu_2 \geq 1$ if $m\geq 2$. Since $|a_1|^2 = \big|\vecbraket{\rho_\mathrm{ss}^{\scriptscriptstyle{K}}}{\rho_\mathrm{ss}^\nodagger}\big|^2$  the lower bound follows immediately.
Although we do not know the singular value $\sigma_2$, the accuracy of the two approximate steady states can nevertheless be compared with only the knowledge of the variance.
In some cases, the second-lowest singular value may also be approximated within the Krylov space.
Notice that it is possible to calculate $\vecbra{\rho_\mathrm{ss}^{\scriptscriptstyle{K}}} \mathcal{\hat L}^\dagger\mathcal{\hat L}^\nodagger \vecket{\rho_\mathrm{ss}^{\scriptscriptstyle{K}}}$ directly inside the Krylov subspace similar to \cref{app:sec:expectations:and:overlaps}.
A non\hyp trivial upper bound can be obtained using $\mu_m/\mu_{D^2}\leq 1$, where $\mu_{D^2}$ denotes the highest eigenvalue. 
Following the same steps as above, this implies
\begin{equation}
\big|\vecbraket{\rho_\mathrm{ss}^{\scriptscriptstyle{K}}}{\rho_\mathrm{ss}^\nodagger}\big|^2 \leq 1 - \frac{\vecbra{\rho_\mathrm{ss}^{\scriptscriptstyle{K}}} \mathcal{\hat L}^\dagger\mathcal{\hat L}^\nodagger \vecket{\rho_\mathrm{ss}^{\scriptscriptstyle{K}}}}{|\sigma_{D^2}|^2} \; .
\end{equation}
Interestingly, it is now possible to quantify the accuracy of the bounds. 
It is governed by the ratio $\sigma_{D^2}/\sigma_2$, which is connected to the condition number of the Lindbladian (on the subspace orthogonal to the steady state).

\begin{figure}[h!]
    \centering
    \subfloat[\label{submat:subfig:conv:evals:a}]{
    \includegraphics[width=0.9\columnwidth]{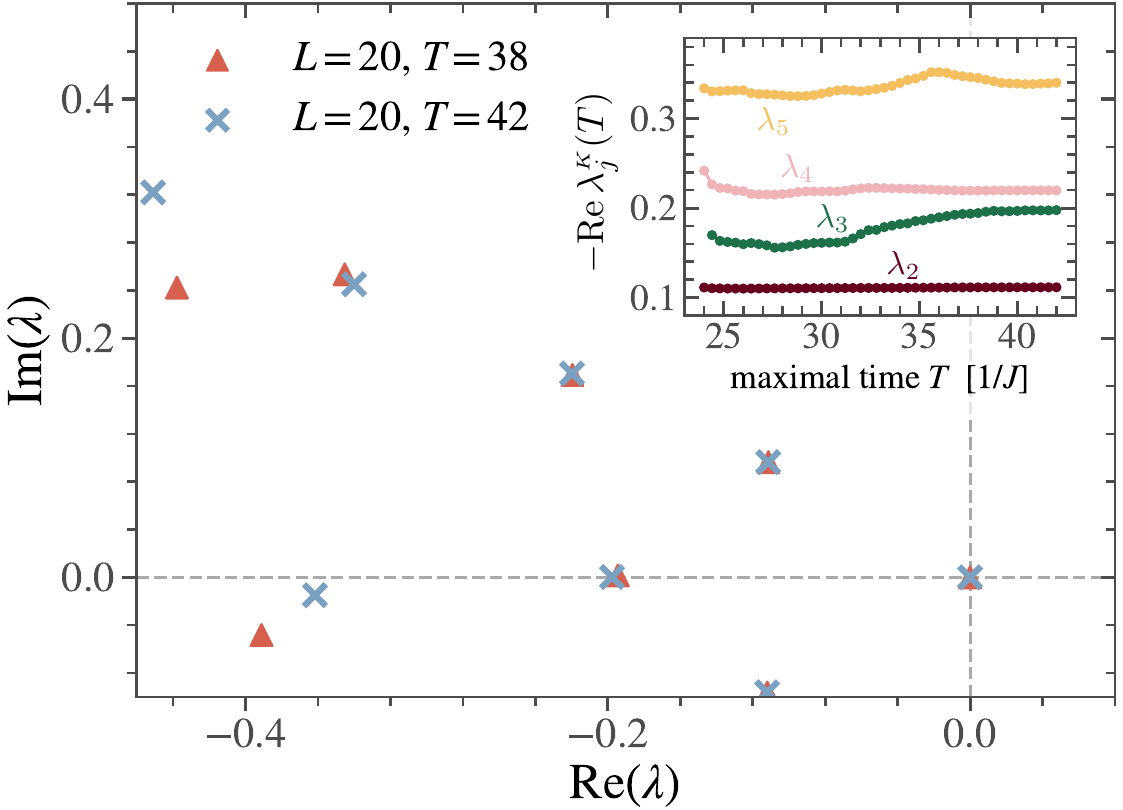}
    }
    \vspace{-
0.3cm}
    \subfloat[\label{submat:subfig:conv:evals:b}]{
    \includegraphics[width=0.92\columnwidth]{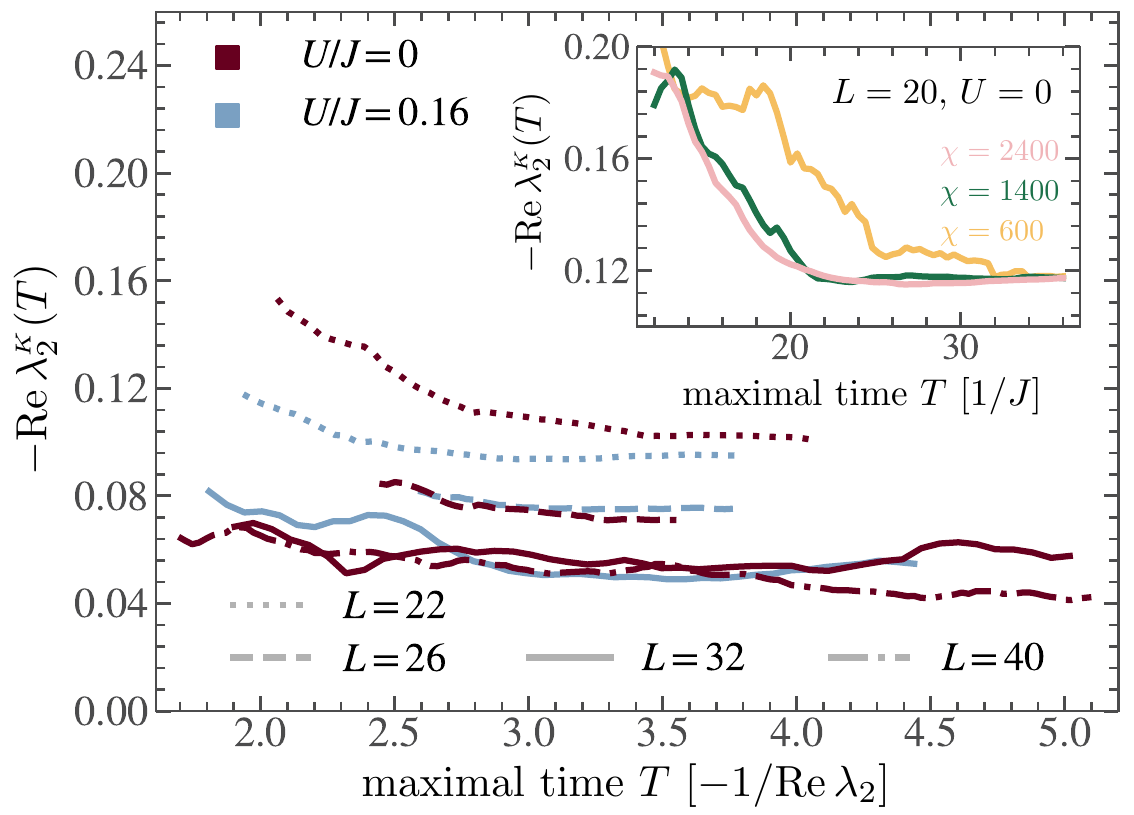}
    }
    \caption{ Convergence analysis for large-scale systems.
    Panel a): \gls{CLIK-MPS} generated spectrum of the Bose Hubbard system with $U/J=0.16$ with $L=20$ at half filling, depending on the maximal evolution time $T$. 
    We fix the Krylov sampling step at $\delta t=0.2$, and evolve to $T=34$ (orange) and $T=38$ (blue), getting Krylov spaces of dimension 340 and 380, respectively.  By varying the maximal time $T$ taken into the Krylov space, we can see if a spectral point is sufficiently converged (see also inset). 
    Inset: Maximal time $T$ dependence of the first four nonzero eigenvalues.
    We set the maximal bond dimension to $\chi=2400$.
    Panel b): Dissipative gap from \gls{CLIK-MPS} for system sizes $L=22, \, 26$ at  $\mathrm \delta t=0.4$ and $L=32$, \updated{$40$} at $\delta t=0.6$ depending on the maximal evolution time $T$ at two different $U/J$. $T$ is given in units of the gap to best display the damping rates the states experience. 
    We set the maximal bond dimension to \updated{$\chi=1600$ for $L=40$ and $\chi=1400$ for all other system sizes}.
    Inset: Dissipative gap depending on $T$ for $L=20, \, U=0$ at half filling with $\mathrm \delta t=0.2$ with three different maximal bond dimensions during time evolution. All eventually converge to the same dissipative gap. Notice that deviations in the intermediate time regime might also originate from the stochastically chosen initial states.
    For all data shown here, we chose $J=1$, $\kappa=2J$, a local dimension $d=N+1$.
    }
    \label{app:fig:convergence:of:spectrum}
\end{figure}
Unfortunately, it is hard to derive error bounds for other parts of the spectrum, but there are other ways to check if a spectral feature is sufficiently converged.
A simple yet effective approach is to compare approximated spectra from different Krylov spaces with each other. 
We suggest varying the maximal evolution time $T$ used to construct the Krylov space, while the sampling interval $\delta t$ remains fixed.
In the following, we refer to the Krylov space with larger $T$ as the \textit{big} Krylov space (having $T_\mathrm{big}$), and to the other as the \textit{small} one ($T_\mathrm{small}$).
This construction ensures that the second Krylov space contains the first, so the \textit{big} Krylov space is expected to yield a more accurate spectrum than the \textit{small} one. 
If $T_\mathrm{big}-T_\mathrm{small}\gg \delta t$, but an eigenvalue is the same in both Krylov spaces, we may say that this eigenvalue is \textit{sufficiently converged}.
Notice that if the Gram matrix $\mathbf M$ and effective Lindbladian $\mathbf L$ have been calculated for the big Krylov space, then the ones in the small Krylov space are trivially known by restricting the matrices to the time-evolved states contained in the small one, and no further overlaps or expectations need to be calculated. This makes convergence analysis particularly easy.

In the insets of \cref{submat:subfig:conv:evals:a} (\cref{fig:spec:random:states}) such an analysis is shown for the dissipative Bose\hyp Hubbard system with 20 sites and 10 particles at $J=1$, $\kappa/J=2$ and $U/J=0.16$ ($U/J=0.5$), respectively.
The spectra obtained by \gls{CLIK-MPS} with complex angles $\alpha=0$ and $\alpha=0.02$, with spacing $\mathrm dt = 0.2/J$ are shown for many different maximal times $T$. While the slowest decaying mode $\lambda_2^{\scriptscriptstyle K}(T)$ is almost constant for $T\geq 25$, indicating sufficient convergence, $\lambda_3$ needs until $T=40$ to converge. Additionally, $\lambda_4$ seems converged out after $T\geq 35$, while $\lambda_5$ would need time evolutions way beyond $T=45$ to converge out. In total, we conclude that maximal time $T=42$ suffices for $U/J=0.16$ and $U/J=0.5$ to resolve both $\Delta_3$ and $\Delta_4$.
In \cref{submat:subfig:conv:evals:a} we furthermore show the whole spectrum for $U/J=0.16$ at identical parameters as in the inset, for two maximal times, underscoring the convergence of the low lying spectrum. 
We conducted the same analysis for all data shown in \cref{fig:spec-and-levecs}, and we chose $T=38$ for $U=0$, $T=42$ for $U=0.16, \, 0.33, \, 0.5$ and lastly $T=44$ for $U=0.66$.
In \cref{submat:subfig:conv:evals:b} we lastly show a part of the convergence analysis for the dissipative gaps provided in \cref{fig:spectrum:dissipative-gap}. We again show the dissipative gap for various system sizes (linestyle) and onsite interactions $U/J $ (color) at half filling, depending on the maximal time $T$ for the Krylov space. 
Since convergence times strongly depend on the system parameters, we express $T$ in orders of the inverse dissipative gap $-1/\mathrm{Re}\, \lambda_2$. We see that all data is sufficiently converged between $T=3.5$ and \updated{$5$} in orders of the inverse dissipative gap. Nevertheless for $L=32$, we evolve further to be sure and finally take the maximal times $T=87$ for $U=0$, $T=81$ for $U=0.16$ and $T=84$ for $U=0.66$ for \gls{CLIK-MPS} and the analysis provided in \cref{fig:spectrum:dissipative-gap}. \updated{For $L=40$ we evolve up to $T=120.0$ with a step $\delta t=0.6$, resulting in $400$ time evolved states.}
\updated{
Error bounds on the steady state approximation quality have been put forward for Krylov methods in the past. For exponential Krylov spaces, they typically depend on the exponential of the dissipative gap $\Delta$, as well as the maximal evolution time $T$ as $\Vert \rho_0\Vert/\vert \vecbraket{l_1}{\rho_0} \vert  \; \mathrm e^{-\Delta T}$, where the prefactor may vary, but the asymptotic behavior is the same \cite{beattie2005convergence}. This however also means, that at least the asymptotic behavior of the error bound is not sharper than normal time evolution methods. 
In our case however, the Krylov space $\mathcal{K}$ is actually a block Krylov space consisting of two independent blocks of Krylov basis vectors, coming from the traceless and traceful initial states.
To understand the impact of the second time evolution on the approximation quality, we need to first consider the basis vectors again.
They are given by $\vecket{\rho_k}, \vecket{\sigma_k}$ and can be written as time evolutions of the initial vectors as
\begin{align}
    & \mathcal{K} = \text{span}\Big\{ \vecket{\rho_k}, \vecket{\sigma_k} \; \text{for} \, k=1, \dots m \, \Big\}\; ,\\
    &\vecket{\rho_k} = \mathrm e^{k\delta t\hat{\mathcal L}} \vecket{\rho_0}\, , \quad \vecket{\sigma_k} = \mathrm e^{k\delta t\hat{\mathcal L}} \vecket{\sigma_0}\, . \nonumber
\end{align}
Written in the right eigenbasis of the Lindbladian, the states are
\begin{equation}
    \vecket{\rho_k} = a_1 \vecket{\rho_\mathrm{ss}} + \sum_{j=2} \mathrm e^{k\delta t \lambda_j} a_j \vecket{r_j}\; , \quad \vecket{\sigma_k} = \sum_{j=2} \mathrm e^{k\delta t (\lambda_j-\lambda_2)} b_j \vecket{r_j} \; ,
\end{equation}
where the coefficients $a_j, \, b_j$ depend on the left eigenvectors.
Since the joint Krylov space contains all linear combinations of these states by definition, we can change the basis of the space and instead look at the states
\begin{equation}
    \vecket{\gamma_k} = \vecket{\rho_k} - \frac{a_2}{b_2} e^{\lambda_2k\delta_t} \vecket{\sigma_k}\, .
\end{equation}
The Krylov space can be rewritten in terms of these states as
\begin{align}
    \mathcal{K} = \text{span}\Big\{ \vecket{\gamma_k}, \vecket{\sigma_k} \; \text{for} \, k=1, \dots m \,  \Big\} ,
\end{align}
where $\vecket{\gamma_0} = \vecket{\rho_0} - \frac{a_2}{b_2} \vecket{\sigma_0}$. Now notice, that the approximation quality of the steady state is better in $\mathcal K$ than in the smaller Krylov space 
\begin{equation}
    \mathcal K_\gamma = \text{span}\Big\{ \vecket{\gamma_k} \; \text{for} \, k=1, \dots m \,\Big\} \, , 
\end{equation}
since it is a subset of $\mathcal K$.
Crucially, this Krylov space contains no contributions from the fastest eigenmode $\vecket{r_2}$, making the convergence speed at least $-\mathrm{Re}\, \lambda_3$. This directly leads to the bound
\begin{equation}
    \Big\Vert \big( \mathds{1} - P_\mathcal{K}\big) \vecket{\rho_\mathrm{ss}}\Big\Vert \leq \frac{\Vert \rho_0\Vert}{a_1} \; \mathrm e^{\lambda_3 m \, \delta t} + \mathcal O\big( e^{(\lambda_3+\lambda_2) m \, \delta t} \big)\, .
\end{equation}
which is exponentially sharper than an errorbound depending on the dissipative gap. Note that we introduced the projector onto the Krylov space, $P_\mathcal{K}$, here.
}
\updated{
By applying the same logic to another Krylov subspace 
\begin{equation}
    \mathcal K_\sigma = \text{span}\Big\{ \vecket{\sigma_k} \; \text{for} \, k=1, \dots m \,\Big\} \, , 
\end{equation}
one gets a similar bound for the slowest decaying eigenvector given by
\begin{equation}
    \Big\Vert \big( \mathds{1} - P_\mathcal{K}\big) \vecket{r_2}\Big\Vert \leq \frac{ \Vert \sigma_0\Vert}{a_2} \; \mathrm e^{(\lambda_3-\lambda_2) m \, \delta t} \, .
\end{equation}
}
\updated{Of course, the complex time step in the second time evolution has some effects on these bounds. Note however that while the complex angle may affect the exponential scalings by a $\mathcal O(\alpha)$ correction (not too severe because $\alpha$ is in general chosen quite small), it most importantly lifts degeneracy of the real part of eigenvalues. This makes it especially important for systems with degenerate slowest decaying modes.}
\section{Numerical Details}
\label{app:sec:numerical:details}
\begin{figure}
\vspace{1em}
\centering
\subfloat[\label{app:fig:rho:MPS}]{
        \includegraphics[width=0.95\columnwidth]{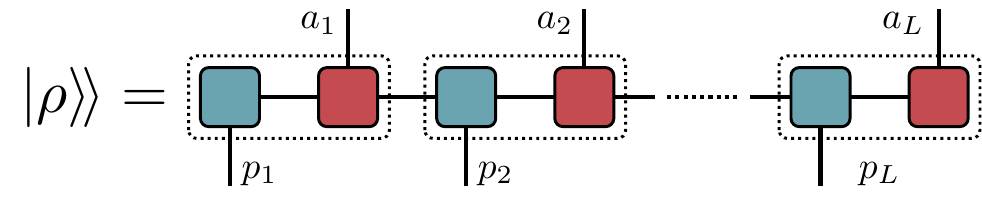}
        }
        \vspace{-0.2cm}
\subfloat[\label{app:fig:L:MPO}]{
        \includegraphics[width=0.94\columnwidth]{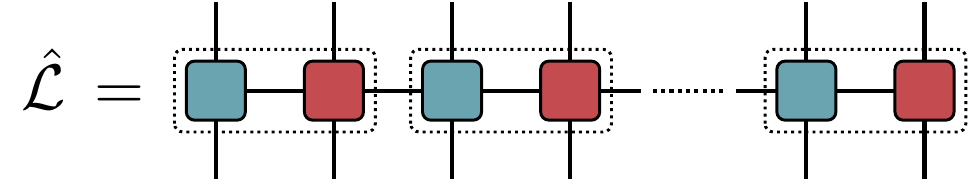}
        }
\caption{The \gls{MPS} representation of a vectorized density matrix (panel a) and the \gls{MPO} representation of a vectorized Lindbladian (panel b) on a doubled lattice. Here $p_j$ labels the $j$\hyp th physical site (blue) and $a_j$ the corresponding auxiliary site (red).}
    \label{app:fig:vectorized:sketch}
\end{figure}

In this section, we present the details of the numerical implementation used for the dissipative time evolution. 
All calculations were performed using the \textsc{SyTen} toolkit \cite{hubig:_syten_toolk,hubig17:_symmet_protec_tensor_networ}.
To employ the \gls{MPS} representation \cite{Schollwoeck2011} in the context of mixed states, we need to vectorize density matrices, which is done by doubling the system size, alternating physical and auxiliary sites, as shown schematically in \cref{app:fig:rho:MPS}. 
In the same way, as shown in \cref{app:fig:L:MPO} we recast the Lindbladian as an operator acting on the vectorized space according to Eq. (2) in the main text \cite{Wolff2020}.
Importantly, alternating physical and auxiliary sites ensures that local Hamiltonians and dissipators remain local in the vectorized description \cite{Casagrande2021}.
In the specific case of the dissipative Bose\hyp Hubbard model considered in the main text, the vectorized Lindbladian features at most next-nearest-neighbor terms.
We utilize the \gls{TDVP} method for \gls{MPS} to simulate dissipative dynamics~\cite{Haegeman2011,Paeckel2019}.
For bosonic systems with a large local physical dimension $d$, as discussed in the main text, the most suitable approach is the \gls{LSE-TDVP} \cite{Yang2020,Grundner2023}.
This method relies on single\hyp site updates in combination with a local subspace expansion \cite{Hubig2015}, offering a speed advantage of a factor of $d$ over the standard \gls{2TDVP}.
Applying \gls{TDVP} to Lindbladian dynamics poses additional difficulties due to the non\hyp Hermitian nature of the generator.
To address this, we adopt a straightforward approach: performing a brute\hyp force Taylor expansion of the exponentials of the local site tensors, which are then used for the time evolution, as explained in \cite{Moroder_diss}.
For all calculations provided in this paper, we chose the timestep $\mathrm{d}t = 0.01$ for time evolutions, \updated{unless otherwise stated}. 
The dissipative Bose\hyp Hubbard model possesses a strong $U(1)$ symmetry associated with the conservation of the total particle number.
In the vectorized representation, this single $U(1)$ symmetry maps onto two $U(1)$ symmetries, corresponding to particle number conservation on each sublattice.
We take advantage of both symmetries within the \gls{MPS} structure to achieve a block\hyp decomposition of the Lindbladian.
The second $U(1)$ symmetry arises because the full vectorized Hilbert space includes unphysical states, and the Lindbladian adopts a block diagonal form that separates physical from unphysical sectors.
\begin{figure}
    \vspace{0.3cm}
    \centering
    \includegraphics[width=0.98\columnwidth]{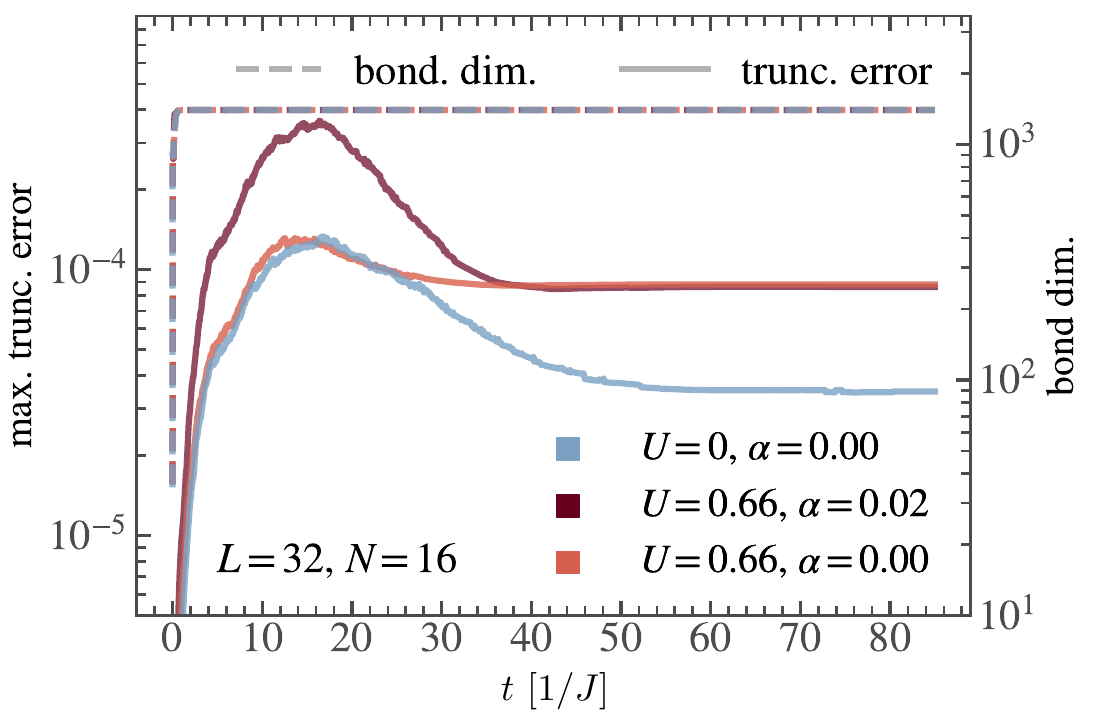}
    \vspace{-0.2cm}
    \caption{Truncation error and right bond dimension during time evolution for large-scale calculations. We consider the Bose\hyp Hubbard model with \updated{32} sites at half filling with $\kappa/J=2$ and $U/J=1$ and $U/J=0$ respectively. We show both the physical time evolution at $\alpha=0$ (blue) for $U=0$ and $U=0.66$ and the complex time evolution with traceless initial state at $\alpha=0.02$ (orange) for $U=0.66$.
    In all cases, the bond dimension increases strongly in the first timesteps up to the maximal bond dimension of $\chi$. Similarly, the truncation error increases in the beginning before decreasing and saturating. 
    The maximal bond dimension $\chi$ is set to $1400$ for all calculations.
    To accurately capture the behavior in the first steps, the time step was initially set to $1\times10^{-3}$, and after 10 steps it was increased to 0.01.
    %
    }
    \label{ap:fig:trunc:error:and:bond:dimension}
\end{figure}
Therefore, explicitly incorporating this symmetry in the implementation is crucial; failing to do so may allow truncation errors to shift weight into the unphysical subspace, potentially introducing unphysical vectors into the Krylov space.
Thanks to the presence of both symmetries, the system can be exactly represented using a local Hilbert space dimension of $d = N+1$. 
Moreover, dissipation suppresses the growth of entanglement in time.
Nonetheless, when fixing a maximal bond dimension $\chi$ for the time evolution, severe truncation errors might occur.
In~\cref{ap:fig:trunc:error:and:bond:dimension} we show the truncation error and bond dimension for one exemplary time evolution. We see that the bond dimension quickly saturates to the maximal bond dimension of $\chi=1400$, while the truncation errors remain moderate. 
Interestingly, low maximal bond dimension $\chi$ seems not to be a problem at all, as we see in the inset of \cref{submat:subfig:conv:evals:b}.
We employ \gls{CLIK-MPS} for a fixed parameter set for three different maximal bond dimensions, and look at the approximated dissipative gap depending on the maximal evolution time. Although at intermediate times, there are large deviations between the different data, they all converge to the same value. 
This is perfectly explained by the effective bond dimension $\chi_\mathrm{eff}$ in the Krylov space (c.f. \cref{eq:app:effective:bdim}), which is the bond dimension that is reachable within the Krylov space. Crucially, besides scaling linearly in $\chi$, it also scales linearly in the Krylov space dimension $D_{\scriptscriptstyle K}$.
Thus, while at mediocre times $\chi=600$ does not have a sufficient $\chi_\mathrm{eff}$ due to low $D_{\scriptscriptstyle K}$, the Krylov space with $\chi=2400$ already has, and it can render the slowest decaying mode correctly. This advantage is diminished by adding more time evolved states to the Krylov space, eventually increasing the effective bond dimension strongly enough to render the eigenmode correctly.
\updated{Note that we found this analysis to be true in some regimes of bond dimension, as shown also in \cref{fig:effective:bond:dim}. However, if the bond dimension during time evolution is chosen much too small, the Krylov space can nevertheless fail in resolving the spectrum accurately. We found this to happen, for instance, when chosing a bond dimension of $\chi=400$ for the same parameters as used in \cref{fig:effective:bond:dim}. Then, the dissipative gap does not converge and shows random wiggles, which is traced back to the worse time evolution.}
To speed up calculations, we can employ parallelization at multiple stages of the framework.
First, the two time evolutions needed in \gls{CLIK-MPS} are independent of each other, so they can be trivially parallelized. 
A similar procedure can be adopted when calculating the overlaps and expectations needed for the effective Lindbladian $\mathbf L$ and Gram matrix $\mathbf G$. 
Here, each overlap and expectation is independent of all the others, and thus the $(N_{\scriptscriptstyle K})^2$ calculations can also be trivially parallelized.
Similarly, expectation values and overlaps for observable calculation in the steady state as described in \cref{app:sec:expectations:and:overlaps} is amenable to parallelization in the same manner.
Regarding truncation of the \gls{MPS} during the dynamics \cite{Paeckel2019}, one has to be careful that truncation errors do not introduce contributions on unphysical states into the Hilbert space.
This problem typically arises if not all the system's symmetries are exploited in the numerics. 
Furthermore, truncation strongly affects the time evolutions with traceless states. 
These states are orthogonal on the steady state sector, which is conserved during time evolution, if there is no truncation.
However, if the state is truncated, in each step a small mass is shifted into the steady state sector, which is enhanced due to the normalization done in each timestep, leading to
\begin{equation}
    \vecbraket{l_1}{\rho(t)} \sim \beta e^{|\mathrm{Re}(\lambda_2)|\,t} \; ,
\end{equation}
where the prefactor $\beta$ is proportional to the truncation error.
The exponential increase originates from the exponential decay of the contribution of the slowest decaying mode, which enhances the weight on the steady state sector during the normalization of $\vecket{\rho(t)}$ at every timestep.
\end{document}